\documentclass{aa}  

\usepackage{graphicx}
\usepackage{txfonts}
\usepackage{subcaption} 
\usepackage{hyperref}
\usepackage[utf8]{inputenc}
\usepackage[english]{babel}
\hypersetup{
    colorlinks=true,
    linkcolor=blue,
    citecolor=blue,
    filecolor=magenta,      
    urlcolor=cyan,
}
\begin{document}

   \title{SPECULOOS - Ultracool Dwarf Transit Survey}
   \subtitle{Target List and Strategy}


   \author{D. Sebastian\inst{1},         
           M. Gillon\inst{1},           
           E. Ducrot\inst{1},         
           F. J. Pozuelos\inst{1,3},    
           L. J. Garcia\inst{1},
           M. N. G\"unther\inst{4},     
           L. Delrez\inst{1,3,5},
           D. Queloz\inst{2},
           B. O. Demory\inst{6},
           A. H.M.J. Triaud\inst{7}, 
           A. Burgasser\inst{8},
           J. de Wit\inst{4},
           A. Burdanov\inst{4},
           G. Dransfield\inst{7},
           E. Jehin\inst{3},
           J. McCormac\inst{9},
           C. A. Murray\inst{2},
           P. Niraula\inst{4},
           P. P. Pedersen\inst{2},
           B. V. Rackham\inst{4}, 
           S. Sohy\inst{3},
           S. Thompson\inst{2},
           V. Van Grootel\inst{3}      
        }

   \institute{Astrobiology Research Unit, University of Li\`ege, All\'ee du 6 ao\^ut, 19, 4000 Li\`ege (Sart-Tilman), Belgium
   \\\email{dsebastian@uliege.be}
         \and
                Cavendish Laboratory, JJ Thomson Avenue, Cambridge CB3 0HE, UK
        \and 
                Space Sciences, Technologies and Astrophysics Research (STAR) Institute, Universit\'e de Li\`ege, All\'ee du 6 Ao\^ut 19C, B-4000 Li\`ege, Belgium         
         \and
                Department of Earth, Atmospheric and Planetary Sciences, MIT, 77 Massachusetts Avenue, Cambridge, MA 02139, USA
        \and
                Observatoire de l’Université de Genève, Chemin des Maillettes 51, Versoix, CH-1290, Switzerland
        \and
                University of Bern, Center for Space and Habitability, Gesellschaftsstrasse 6, 3012 Bern, Switzerland
        \and
                School of Physics \& Astronomy, University of Birmingham, Edgbaston, Birmingham B15 2TT, United Kingdom
        \and 
                University of California San Diego, 9500 Gilman Drive, La Jolla, CA 92093
        \and
                Department of Physics, University of Warwick, Coventry, CV4 7AL
             }
   \titlerunning{The SPECULOOS project}\authorrunning{Sebastian et al.}
   \date{Received ; accepted , }

 
\abstract
  {One of the most promising avenues for the detailed study of temperate Earth-sized exoplanets is the detection of such planets in transit in front of stars small and nearby enough to make possible their thorough atmospheric characterisation with next generation telescopes like the James Webb Space telescope (JWST) or Extremely Large Telescope (ELT). In this context, the TRAPPIST-1 planets form an unique benchmark system that has gathered the interest of a large scientific community.
  }
  {The SPECULOOS survey is an exoplanet transit survey, that targets a volume-limited (40\,pc) sample of ultracool dwarf stars (spectral type M7 and later). It is based on a network of robotic 1\,m telescopes especially designed for this survey, and its strategy for its brighter and earlier targets leverages on the synergy with the ongoing space-based exoplanet transit survey TESS.
  }
  {We define the SPECULOOS target list as the sum of three non-overlapping sub-programs incorporating the latest type objects ($T_{\rm eff}\lesssim$ 3000\,K): Program~1: 365 dwarfs that are small and nearby enough to make possible the detailed atmospheric characterisation of an `Earth-like' planet with the upcoming JWST, Program~2: 171 dwarfs of M5-type and later for which a significant detection of a planet similar to TRAPPIST-1b should be within reach of TESS, and Program~3: 1121 dwarfs later than M6-type that aims to perform a statistical census of short-period planets around ultracool dwarf stars.
  }
  {Our compound target list includes 1657 photometrically classified late-type dwarfs. 260 of these targets are classified for the first time as possible nearby ultracool dwarf stars. Our general observational strategy is to monitor each target for 100 to 200\,hr with our telescope network, by efficiently using the synergy with TESS for our Program~2 and a fraction of the targets of Program~1.
  }
  {Based on Monte Carlo simulations, we expect to detect up to a few dozens temperate, rocky planets, a handful of them being amenable for atmospheric characterisation with JWST and other future giant telescopes which will improve drastically our understanding of the planetary population of the latest-type stars.
  }

   \keywords{Planetary systems  --
               Stars: low-mass --
               Catalogs --
               Astrobiology
               }

   \maketitle
%
\section{Introduction}\label{section:intro}
Thousands of transiting exoplanets have been detected in the past years using ground- and space-based observatories. For some of them, their eclipsing configuration not only allows us to measure directly their sizes, but also to probe a variety of physical properties, ranging from a full orbital characterisation to their atmospheric characteristics. 
In the past decade, atmospheric studies of giant planets have provided constraints on exospheres, molecular and atomic species, elemental ratios, temperature profiles, clouds and even atmospheric circulations (see \cite{Madhusudhan19} for a review). The field aims now to perform similar atmospheric studies for smaller and more temperate planets (like GJ 1214b \cite{cha09}), with the ultimate goal of probing the atmospheric composition of potentially habitable rocky planets. An important step in this direction was achieved recently by \citet{tsiaras2019} and \citet{benneke19} who independently showed the existence of water vapor in the atmosphere of the temperate "mini-Neptune" K2-18 b.

Unfortunately, with current and upcoming instrumentation, such studies are far beyond reach for rocky planets orbiting solar-type stars. Nevertheless, the signal-to-noise ratio (SNR) for eclipse spectroscopy measurements increases for smaller and cooler dwarf stars, while their lower luminosities result in more frequent planetary transits for the same stellar irradiation. As a matter of fact, several studies (e.g. \citealt{Lisa2009,dewit2013,Morley2017}) have shown that the upcoming James Webb Space Telescope (JWST) should be able to probe the atmospheric composition of a potentially habitable Earth-sized planet, but only if it transits a star that is both very nearby (<~ 15\,pc) and very small (<~ 0.15\,$R_{\odot}$), with spectral type M6 or later.

Not many late-type M-dwarf stars that harbour transiting, rocky planets are known so far. The first detection was Kepler-42 (spectral type ~M4) with its compact system of sub-Earth-sized planets \citep{Muirhead12}. It was also a first indication that short-period rocky planets could be more common around low mass stars than for solar-type stars \citep{howard12,Ullman19}.
The detection of seven transiting planets orbiting TRAPPIST-1 \citep{Gillon2016,Gillon2017,Luger2017a}, a close by (12\,pc) M8\,V star, delivered the up-to-date best targets for atmospheric characterisation with the upcoming JWST in the temperate Earth-sized regime \citep{LustigYaeger2019,macdonald19}. Other transiting rocky planets were detected around mid-to-late-type M-dwarfs by MEarth (GJ1132 (M4\,V, \citealt{BertaT15}); LHS 1140 (M4.5\,V, \citealt{dittmann17,ment2019})), and by TESS (LP 791-18 (M6\,V, \citealt{crossfield19}); LHS 3844 (M6\,V, \citealt{Vanderspek2019})).

Despite these few detections, not much is known about the structure of planetary systems of late-type M-dwarfs. This is especially true for ultracool dwarfs. The classical definition of ultracool dwarfs (hereafter UCDs) includes dwarfs with spectral type M7 and later as well as brown dwarfs (BDs) \citealt{kirkpatrick97};\cite{Kirkpatrick05}).
Surveys like MEarth \citep{Nutzman2008} and TESS \citep{Ricker2015}, which use relatively small (< 50\,cm) telescopes, have a high detection potential for mid-type M-dwarfs (M3 to M5), but this potential drops sharply for later-type objects \citep{Sullivan2015,barclay18}. Up to now, TRAPPIST-1 remains the only transiting system discovered around an UCD, while some statistical constraints could be inferred from null results of several projects \citep{Demory2016,he17,segear19,lienhard20}.

Several theoretical studies have tried to predict which kind of planetary systems could be formed around UCDs (\cite{payne07,raymond07,lissauer07,montgomery09,alibert17,coleman19,schoonenberg19,miguel20}). They predict a variety of outcomes, from water rich to water poor planets and various orbital architectures. Notably, \cite{coleman19} inferred the period and mass distribution of planets for typical UCDs. They found that most systems around UCDs should have commonly one or more planets with a period distribution that peaks for short period planets. One the other hand, \cite{miguel20} found a bimodal distribution for the distance of the planets; one group with very short periods and another group with larger periods, located beyond the snow line. observations will help to distinguish between these various models.

Our current lack of knowledge about the planetary population of UCDs -- combined with the opportunity that they represent for the atmospheric characterisation of temperate rocky planets -- made a strong case for the development of a photometric survey specifically designed to explore the nearest UCDs for transits of planets as small as the Earth, or even smaller. 
In this context, we developed a project, SPECULOOS (Search for habitable Planets EClipsing ULtra-cOOl Stars). Its main goals are (1) search for rocky planets, well-suited for atmospheric characterisation with future facilities like JWST and (2) more globally, to perform a volume-limited (40pc) transit search of UCDs to achieve a statistical census of their short-period planet population \citep{Delrez2018b,Gillon2018}.

Given the relative faintness of those objects and the lessons learned from the TRAPPIST-UCDTS prototype survey (\citealt{Gillon2013,Burdanov2018,lienhard20}), we concluded that a ground-based network of 1\,m class, robotic telescopes should have a good detection efficiency for planets as small as the Earth -- and even smaller -- for a large fraction of UCDs within 40pc.

Based on these considerations, we developed the SPECULOOS robotic telescope network. It is composed of the SPECULOOS Southern Observatory (SSO) with four telescopes at ESO Paranal Observatory (Chile) \citep{Delrez2018b,Jehin2018,murray20}, the SPECULOOS Northern Observatory (SNO) with one telescope at Teide Observatory in Tenerife (Canary Islands) \citep{Delrez2018b}, as well as SAINT-EX (Search And characterIsatioN of Transiting EXoplanets, \citep{demory2020}) with one telescope in San Pedro M\'{a}rtir observatory (Mexico).
In addition, the two 60\,cm TRAPPIST robotic telescopes of the University of Liège (one in Chile, the other in Morocco; \citep{Gillon11,jehin11}), while not being officially parts of the SPECULOOS network, have devoted a fraction of their time to support the project, focusing on its brightest targets. SSO, SNO and SAINT-EX use 1\,m equatorial telescopes, equipped with identical NIR sensitive CCD cameras. All facilities became fully operational in 2019.

The first detection from this network was the detection of the binary brown dwarf 2MASSW J1510478-281817, a nearby brown dwarf binary with a tertiary brown dwarf companion. It is one of only two double-lined, eclipsing brown dwarf binaries known today and, thus, a rare benchmark object to constrain the masses, radii and ages of such objects \citep{triaud20}.

Up to now, a volume-limited list of UCDs within 40\,pc does not exist. In the past decades, large scale photometric, astrometric, and spectroscopic surveys based on SDSS \citep{hawley02}, 2MASS \citep{cruz03}, WISE \citep{kirkpatrick11}, or Gaia \citep{reyle18,smart19} led to the detection of most UCDs, known to date. \cite{Gagliuffi2019} estimated that the current census within 25\,pc is incomplete with only 62\% and 83\% for late-M and L dwarfs known respectively. Thus it is not surprising, that some nearby objects are still regularly found (e.g. Scholz's star (\cite{scholz142}).

To build the SPECULOOS target list, we decided to start from the Gaia DR2 point source catalogue \citep{Gaia2016, Gaia2018} and to cross-match it with the 2MASS point-source catalogue \citep{2MASS}, enforcing the agreement between the two catalogues not only in terms of position but also in terms of effective temperatures inferred from different photometric indicators. This selection procedure is presented in Section 2. Its main by-product is a catalogue of M and L dwarfs within 40\,pc (appended to this paper in electronic form).

The rest of the paper is structured as follows. In Section 3, we describe how we divided the SPECULOOS target list into three non-overlapping sub-samples, each corresponding to three different scientific programs. In Section 4, we describe the properties of the whole SPECULOOS sample, notably how our photometrically-derived spectral types compare to spectroscopically-derived spectral types from the literature. In Section 5, we describe in detail the survey observational strategy, based notably on reachable phase coverage and the possible input of TESS observations. In Section 6, we introduce the SPECULOOS target scheduler. Section 7 presents the detection potential of the project as derived from Monte Carlo simulations. Finally, we give our conclusions in Section 8.

\section{Target selection}\label{section:target_sel}
To build our target list, we first developed a catalogue of M- and L-dwarfs within 40\,pc,
starting with the 35,781 objects in the Gaia DR2 catalogue with a trigonometric parallax $d \ge 25$\,mas. 
For each of them, we (1) applied the correction to the Gaia DR2 parallax recommended by \cite{Stassun2018}; 
(2) computed the J2000 equatorial coordinates considering only the proper motion 
as measured by Gaia (the epoch of Gaia DR2 coordinates is J2015.5); 
(3) computed the absolute  magnitude $M_G$ from the apparent $G$-band magnitude and the Gaia distance modulus measured; 
(4) computed an estimate of the effective temperature $T_{\rm eff}$ based on the empirical law $T_{\rm eff}(M_G)$ 
of \citet{PM2013} (hereafter PM2013) and assuming a systematic error of 150\,K added quadratically to the error propagated from the error on $M_G$. 
from the resulting list we discarded all objects with $M_G < 6.5$ or a colour $G_{BP} - G_{RP} < 1.5$ to keep only dwarf stars later than $\sim$K9-type. We also discarded objects missing a $G_{BP} - G_{RP}$ colour in Gaia DR2 and ended up with 21,137 potential nearby M- and L-dwarfs.

\begin{figure*}[h!]
    \centering
    \includegraphics[width =\columnwidth]{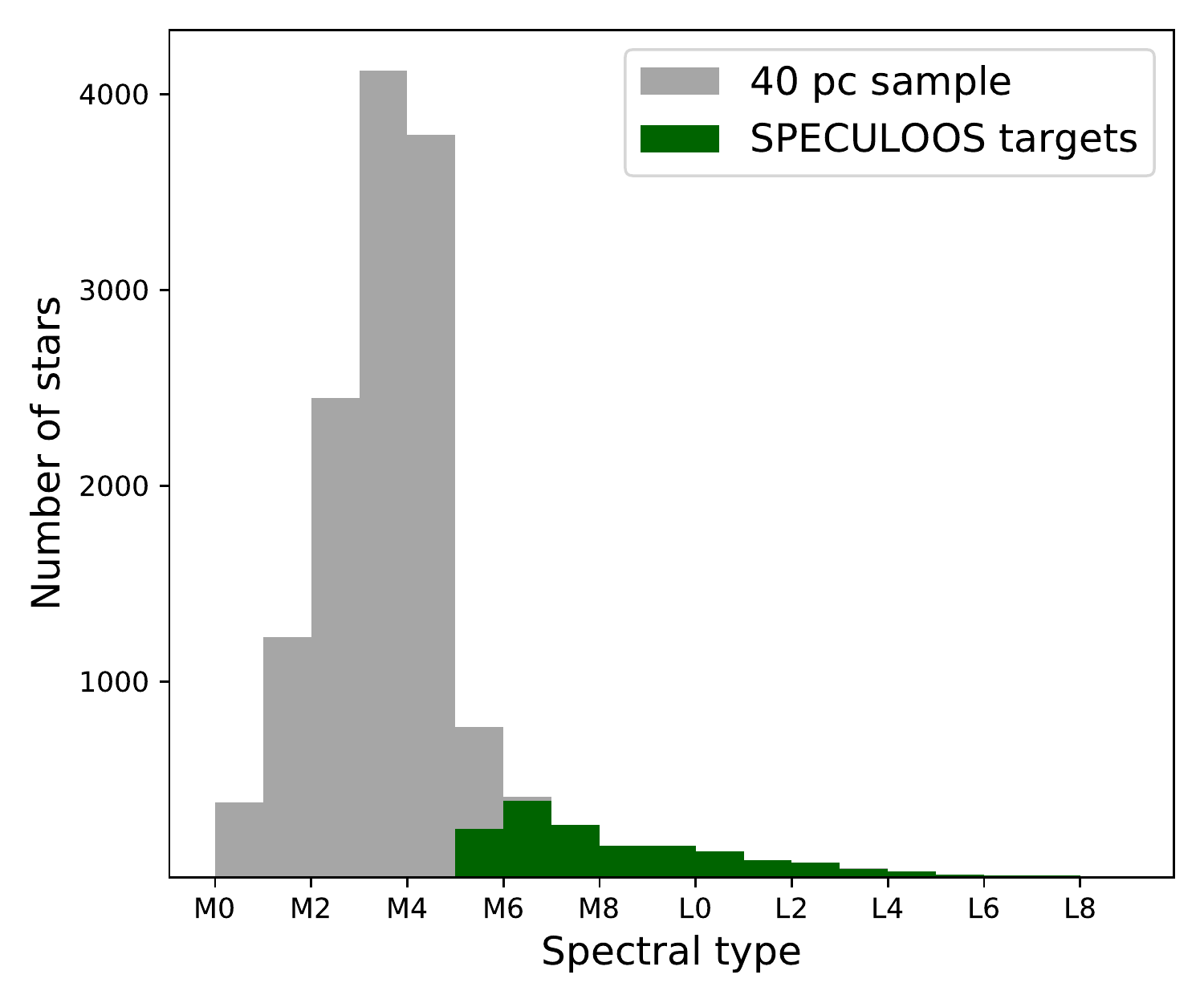}
    \includegraphics[width =\columnwidth]{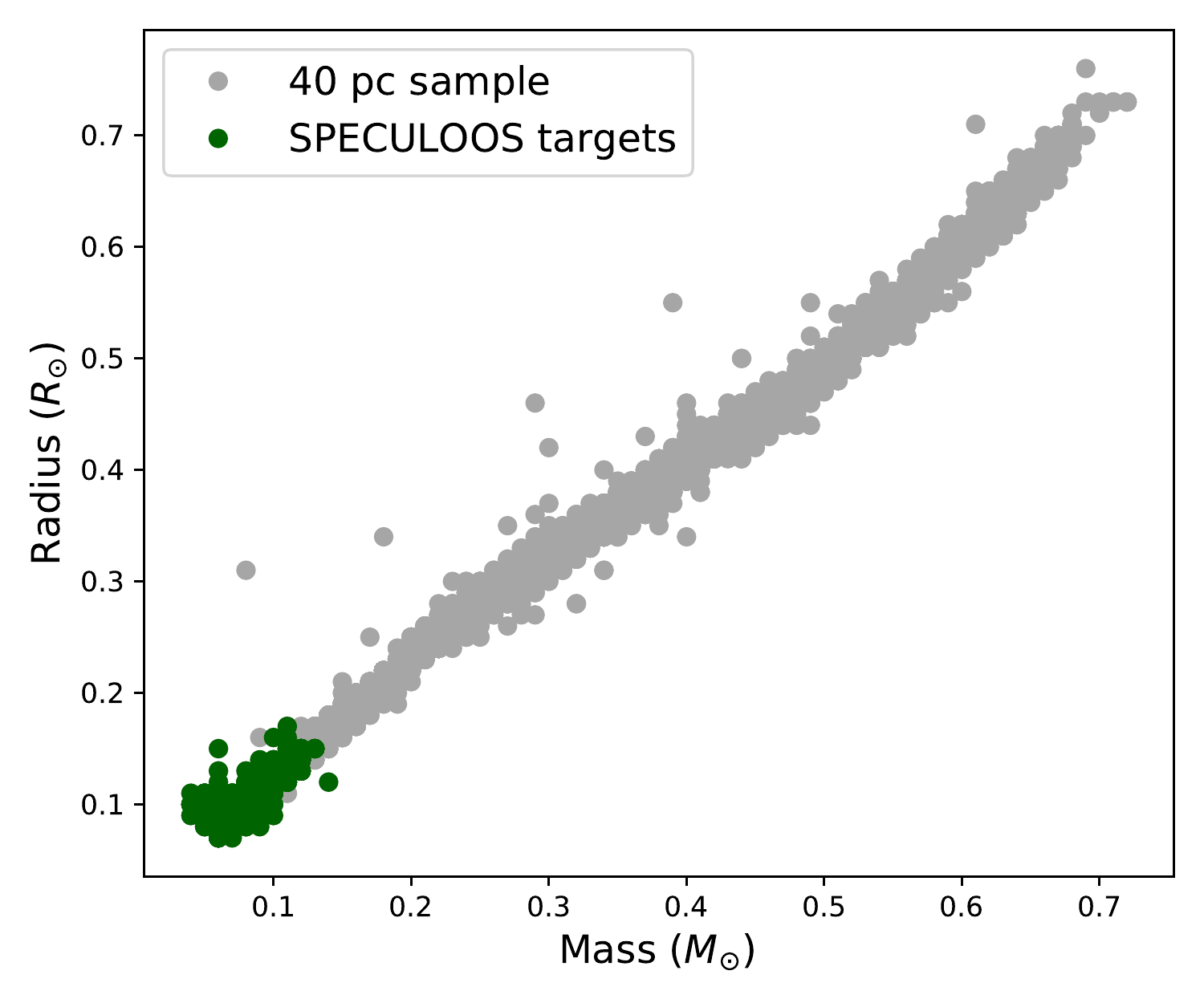}
    \caption{Spectral type distribution ($left$) and mass-radius diagram ($right$) for our 40pc ML-dwarfs catalogue. Gray: The Gaia-2MASS 40\,pc sample for late-type stars, Green: The SPECULOOS targets (See section \ref{section:programs}).}
    \label{fig:cat}
\end{figure*}

Afterwards, we cross-matched each of these objects with the 2MASS point sources within $2'/d$ (= $3''$ at 40\,pc). This $1/d$ dependency of the search radius aimed to take into account that, for the nearest stars, the astrometric position at a given time can differ significantly from the one computed by correcting the J2015.5 position from the proper motion. This is due to the significant amplitude of their 3D motion, meaning that their radial velocity should also be considered. For instance, there is an astrometric difference $>$30'' between the J2000 position of Proxima Centauri as measured by 2MASS and the one computed from its Gaia DR2 J2015.5 coordinates and proper motion.

For each 2MASS object falling within 2'/$d$ of a selected Gaia DR2 object, we computed two estimates of $T_{\rm eff}$, one based on the $T_{\rm eff}(G-H)$ empirical relationship of PM2013 assuming a systematic error of 150K, and one based on the $T_{\rm eff}(M_H)$  empirical relationship of \citet{F2015} (hereafter F2015), assuming a systematic error of 100K, and with the absolute magnitude $M_H$ computed from the apparent $H$-band magnitude measured by 2MASS and from the distance modulus measured by Gaia. To extend the empirical relationship of F2015 for late-type M-dwarfs to earlier-type stars, we derive $T_{\rm eff}(M_H)$ for targets with $M_H < 7.83$ using the empirical relationship of PM2013. This value of 7.83 was found to ensure the continuity of the two laws. We then computed for each Gaia - 2MASS couple the following metric: \\
\begin{equation}
\begin{array}{l}
\left ( \frac{T_{\rm eff}(M_G)-<T_{\rm eff}>}{\sigma_{T_{\rm eff}(M_G)}} \right )^2 + 
\left ( \frac{T_{\rm eff}(G-H)-<T_{\rm eff}>}{\sigma_{T_{\rm eff}(G-H)}} \right )^2 + \\
\left ( \frac{T_{\rm eff}(M_H)-<T_{\rm eff}>}{\sigma_{T_{\rm eff}(M_H)}} \right )^2 + 
\left (  \frac{\delta_{position}}{\sigma_{\delta_{position}}}\right )^2
\end{array}
\end{equation} where $<T_{\rm eff}>$ is the mean of the three temperature estimates and where
\begin{eqnarray} 
\delta_{position} &=& \arccos{(\sin{\delta_1}\sin{\delta_2} } \\
&+& \cos{\delta_1}\cos{\delta_2}\cos{(\alpha_1 - \alpha_2)}), 
\end{eqnarray} $\alpha_i$ and $\delta_i$ being the right 
ascension and declination of the star $i$, and where $\sigma_{\delta_{position}}$ is the error on the position difference between the two objects computed from propagation of the errors on $\alpha$ and $\delta$ quadratically (assuming a measurement error of 1.21'' for 2MASS, \cite{Stassun2018}),  summed quadratically to an error of 85''/$d$ to take into account the significant 3D motion of the nearest stars. This value of 85'' is the Gaia-vs-2MASS position difference of SCR1845-6357, the very nearby (4\,pc) star with the highest  $\delta_{position}$ / distance ratio.

Each Gaia DR2 object was cross-matched with the 2MASS object within 2'/$d$ that minimized its metric function, i.e. with the nearest position and the best match in terms of $T_{\rm eff}$ as derived from $M_{G}$, $M_{H}$ and $G-H$. For 3,660 objects, no cross-match was found, i.e. no 2MASS object was found within 2'/$d$. For the remaining 17,477 objects (21,137 - 3,660), the three temperature estimates $T_{\rm eff}(M_G)$, $T_{\rm eff}(G-H)$, and $T_{\rm eff}(M_H)$ were compared between each other. In case of discrepancy at more than $2\sigma$, the object was discarded. 2,848 objects were rejected through this comparison, leaving 14,629 objects.
For these objects, we (1) derived an estimate of the spectral type ($\rm SpT$) by inverting the empirical relationship $T_{\rm eff}(M_H)$ of F2015, assuming an internal error of 113K for it; 
(2) computed an estimate of the $I_c$-band magnitude from the 2MASS $J$-band magnitude and the spectral type  estimate using online tables with empirical colours as a function of spectral type\footnote{\url{http://www.stsci.edu/~inr/intrins.html}}; 
(3) computed a $J$-band bolometric correction $BC_{J}$\footnote{We used the $BC_J(\rm SpT)$ relationship of F2015 (assumed internal error = 0.163), and for stars earlier than SpT M6.64 (selected so to ensure the continuity of the two laws), we used the $BC_J(T_{\rm eff})$ relationship of PM2013 assuming an internal error of 0.2.}; (4) computed an estimate of the bolometric luminosity $L_{\rm bol}$ (+ error) from $M_J$ and $BC_J$; 
(5) computed an estimate of the radius $R_\star$ (+ error) from $L_{\rm bol}$, $T_{\rm eff}$, and the relationship $L_{\rm bol} = \sigma 4 \pi R_\star^2 T_{\rm eff}^4$, where $\sigma$ is the Stefan-Boltzmann constant; (6) computed an estimate of the mass $M_\star$ (+ error) from the empirical relationship of \cite{Mann2019} (assumed internal error = 3\%) for objects earlier than L2.5, assuming them to be low-mass stars \citep{Dieterich2014}. 
For later objects (i.e. brown dwarfs), we assumed the following relationship to derive a crude estimate of the mass: $M_\star = 0.075 - (\rm SpT-12)\times 0.0005 M_\odot$. As the spectral type of a brown dwarf is not only correlated with its mass but also its age, this relationship has no ambition to be accurate at all and should not be used as scientific reference for these targets. It just aims to represent that, statistically speaking, a hotter brown dwarf tends to be more massive than a colder one. We assign an error of 80\% to the mass as these brown dwarfs cannot have masses substantially larger than 0.075\,$M_\odot$ but could be as low as a few ten Jupiter masses. 
At that stage, we rejected another batch of 520 objects for which at least one of the following conditions was met: 
\begin{itemize}
    \item The computed radius was smaller than 0.07 $R_\odot$, i.e. too small for an ultracool dwarf \citep{Dieterich2014}.
    \item The number of 2MASS objects within 2' was over 250 and the $K$ magnitude larger than 12.5,  making likely a confusion case (galactic disk + bulge).
    \item The inferred spectral type  was later than M5.5 [M9], $K$ was larger than 8 (so no saturation in 2MASS images), 
    and still the $J - K$ colour was smaller than 0.6 [1.0], suggesting a wrong cross-match or a confusion case.
    \item The inferred mass was smaller than 0.2 $M_\odot$ while the inferred radius was larger than 0.4 $R_\odot$, 
    i.e. too large for a low-mass M-dwarf.
\end{itemize}
We ended up with a 40\,pc M+L dwarfs catalogue containing 14,109 objects. Still, we noticed that some well-known nearby late-M and L-dwarfs were missing from the catalogue because they had no parallax 
in Gaia DR2, including very nearby objects like Scholz's star (M9.5+T5.5 at 6.7\,pc, \citealt{Burgasser2015}), Luhman-16 (L7.5+T0.5 at 2.0\,pc, \citealt{Luhman2013}), or Wolf 359 (M6.0 at 2.4\,pc, \citealt{Henry2004}). 
To account for this, we cross-matched our catalogue with the spectroscopically verified sample of M7-L5 classical ultracool dwarfs within 25\,pc of \cite{Gagliuffi2019}. For each object within this sample and not present in our catalogue, we (1) used  the $T_{\rm eff}(\rm SpT)$ empirical relationship of F2015 to estimate the effective temperature, (2) used the same procedure than for the Gaia DR2 objects to estimate the bolometric correction, the luminosity, the radius, and the mass. 
We discarded objects flagged as close binary in the catalogue of \cite{Gagliuffi2019}, objects with an inferred size smaller than 0.07 $R_\odot$, objects with an inferred mass greater than 0.125 $M_\odot$ (too massive for an ultracool dwarf, suggesting a blend or a close binary case), and objects later than M9.0 with a $J-K$ colour index
smaller than 1.0 and a $K$ magnitude larger than 8. 
This procedure added 59 objects to our catalogue, for a total of 14,168. Except for one target, 2MASS J21321145+1341584 which does not have a corresponding source in Gaia DR2, we cross-matched all added objects with the Gaia DR2 catalogue.
Figure \ref{fig:cat} shows the spectral type distribution and the mass-radius diagram of the 14,168 objects. One can notice that our catalogue of 40\,pc M + L dwarfs\footnote{
The full catalogue is only available as 40pc\_list in electronic form at the CDS via anonymous ftp to \url{cdsarc.u-strasbg.fr} (130.79.128.5) or via \url{http://cdsweb.u-strasbg.fr/cgi-bin/qcat?J/A+A} }. is dominated by $\sim$M4-type objects, in line with earlier results (e.g. \citealt{Henry2018}).

\begin{figure}
    \centering
    \includegraphics[width=\columnwidth]{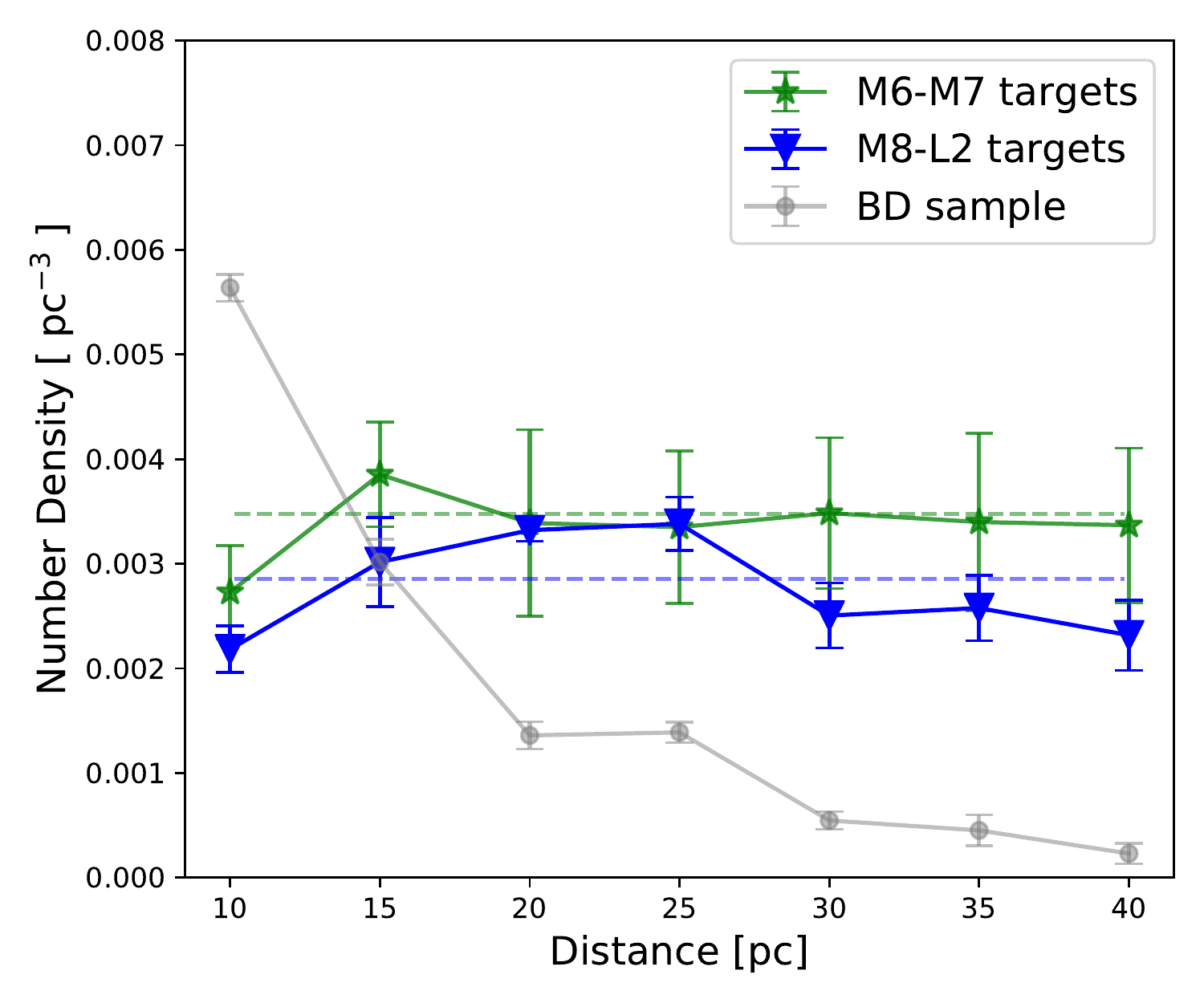}
    \caption{Number densities per distance for different spectral types. Green asterisks: sub-types M6 and M7; blue triangles: sub-types M8 to L2; and gray points BDs with types later than L2. The horizontal lines mark the average densities.
    }
    \label{fig:dense}
\end{figure}

Our final goal was to build the target list of SPECULOOS, using a consistent target selection method, rather to present a complete sample of late type stars. Therefore, we did not try to recover stars earlier than M6 absent from our catalogue because they do not have a Gaia DR2 parallax. Extrapolating our results for ultracool dwarfs to earlier M-dwarfs, we estimate that there must be at most a few hundreds of them. 
To estimate the number of stars that were still missed by our selection process, we derived the number density of stars per cubic parsec. For the early M-dwarf sample (M0 to M5), we derive a distance independent number density of (46.6$\pm$0.8)\,x\,10$^{-3}$\,pc$^{-3}$. Figure \ref{fig:dense} shows the number density for the late M-dwarfs and BDs of our 40\,pc sample. We see a lower density for targets closer than 15\,pc. This is mainly due to a few nearby stars that are not covered by the \cite{Gagliuffi2019} survey and have missing parallaxes in Gaia DR2.
For M6 to M7 stars, we derive a distant independent number density between 10 and 40\,pc of (3.4$\pm$0.2)\,x\,10$^{-3}$\,pc$^{-3}$. Assuming a homogeneous distribution, we conclude that this sample, as well as our early M-dwarf sample are distance limited with no or negligible brightness selection.
For late UCDs with spectral types M8 to L2, we see a drop for targets between 30 and 40\,pc, which can be explained by a selection effect that leads to a lack of stars in the order of $10^{2}$, mostly due to their faintness or crowding in the galactic plane. Finally for BDs with spectral types later than L2, we see a steep drop with distance. The number density of BDs in our sample decreases by 50\% between 10 and 15\,pc. This lack of BDs is expected due to the intrinsic faintness of those objects and the limiting magnitude of Gaia \citep{reyle18,smart19}.

Within 25\,pc, we find 214 targets with a photometric spectral type between M6.5 to L0, corresponding to a mean number density of (3.2$\pm$0.3)\,x\,10$^{-3}$\,pc$^{-3}$. This value is about 25\% lower than the raw volume-corrected value derived by \cite{Gagliuffi2019} ((4.1$\pm$0.3)\,x\,10$^{-3}$\,pc$^{-3}$). This absolute difference can be explained by our selection method, which excludes (i) most close binaries and (ii) blended stars.

\section{Observing Programs}\label{section:programs}
As described in Sec. \ref{section:intro}, the core science cases of the SPECULOOS survey can be broken down into (1) the search for transiting, rocky planets well-suited for atmospheric characterisation with future facilities, and (2) a statistical census of temperate planets around UCDs. To optimise these goals, we divide the survey into three non-overlapping programs.

Anticipating the launch of JWST\footnote{\url{https://www.jwst.nasa.gov/content/about/faqs/facts.html}}, we focus in our first program on a census of targets for which the atmospheric properties of an `Earth-like' planet could be studied in some detail by an ambitious JWST transit spectroscopy observation \citep{gillon20}. As `Earth-like', we denote a planet with the same mass, same size, same atmospheric composition and same irradiation from the host star than the Earth. 
Our second program focuses on a census of temperate rocky planets in the overlapping region between SPECULOOS and TESS, in the M5-M6 spectral range. As temperate, we denote a planet that receives four times the irradiation the Earth receives from the Sun, like TRAPPIST-1b. This irradiation is as a boundary, with equilibrium temperatures low enough to allow habitable regions on such tidally locked planets \cite{menou13}.
In this spectral type range, TESS has a declining sensitivity to Earth-sized planets due to a lack of photons, while SPECULOOS detection efficiency is decreased by the larger periods of temperate planets (relative to later M-dwarfs) that extends the telescope time required per target. A synergetic approach combining the long and continuous observations of TESS and the higher photometric precision of SPECULOOS could thus make easier detections that would be difficult to achieve individually for both surveys. 
Our third program focuses on a statistical search for transiting exoplanets within all remaining targets. The criteria to select the targets of the three programs from our 40\,pc list are presented in the following subsections.

\subsection{Program~1: `Earth-like' planets for JWST} \label{section:program1}

To set up the target list of our first program, we computed the typical signal-to-noise ratio (SNR) achievable in transit transmission spectroscopy by a 200\,hr observation with JWST/NIRSPEC for all 14,168 objects within our 40\,pc sample. Our aim here is to efficiently probe the atmospheric properties of an `Earth-like' planet.
For the typical amplitude of the transmission signals, we used the following equation \citep{Winn2010}:
\begin{equation}
\Delta \delta = 2 N_H \delta \left( \frac{H}{R_\oplus} \right),
\end{equation} 
where $H$ is the atmospheric scale height. to derive $H$ we assume an `Earth-like' planet with the same atmospheric composition like the Earth, an isothermal atmosphere with a mean molecular mass of 29\,amu, and a temperature equal to the equilibrium temperature of the planet (with an Bond albedo of 0.3 and irradiation of $1\,S_{\oplus}$). $\delta$ is the transit depth, and $R_\oplus$ is the Earth's radius. We assumed a value of 5 for $N_H$, the number of scale heights corresponding to a strong molecular transition. For the assumed planets, the orbital distance corresponding to an `Earth-like' irradiation was computed based on the stellar luminosity, the corresponding 
orbital period was computed using Kepler's third law combined with the stellar mass estimate, and the duration of a central transit was computed using equation 15 from \cite{Winn2010}. 
Then, we computed the JWST/NIRSPEC  (Prism mode) noise at 2.2 $\mu$m for a spectral bin of 100\,nm and for an exposure sequence with the same duration as the transit using the online tool PandExo \citep{Batalha2017}. We assumed a red noise over a  transit timescale of 30\,ppm that we added quadratically to the white noise estimate of PandExo. We computed the number of transits observed within the 200\,hr JWST program, assuming for each transit observation a duration equal to the transit duration 
plus 2.5\,hr (for pointing, acquisition, plus out-of-transit observation). The noise per transit was then divided by the square root of the number of observed transits, and we added quadratically to the result an absolute floor noise of 10\,ppm. 
At the end, we obtained for each target a transmission signal-to-noise (SNR) by dividing the transmission amplitude by the global noise. Figure \ref{fig:jwst} (left) shows the resulting $\rm SpT-SNR$ distribution. In this Figure, we draw a line at $\rm SNR$ = 4, assuming this value to be an absolute minimum for deriving meaningful constraints on the atmospheric composition of our assumed `Earth-like' planets. TRAPPIST-1 is shown as a magenta star symbol in the Figure. For the majority of the targets in the whole 40\,pc sample, the achievable SNR is $\sim 1$. 
 
We find that 366 objects -- including TRAPPIST-1 -- have a SNR $\ge$ 4, and only 44 of them have a SNR larger than TRAPPIST-1. These 366 objects constitute the target list of SPECULOOS Program~1. Of the 366, 92 have a spectral type earlier than M6 (and none of them is earlier than M4). We choose to also include these earlier targets in our target list even if they are not {\it bona fide} UCDs.

\subsection{Program~2: Temperate, Rocky Planets from TESS}\label{section:program2}

We set up the targets for our second program identifying the synergetic region between SPECULOOS and TESS within our 40\,pc sample. For this, we derived first the detection threshold for a temperate, Earth-sized planet for both surveys. In here, we assumed a planet still to be temperate when it receives four times the irradiation the Earth receives from the Sun.  

First, we estimated the photometric precision that should be achieved by TESS within 30\,min for each object. For this, we adopted the moving 10th percentile of the TESS RMS in one hour, as function of the T magnitude, published in Figure 6 of the TESS Data Release Notes NASA/TM-2018-000000\footnote{\url{https://tasoc.dk/docs/release\textunderscore notes/tess\textunderscore sector\textunderscore04\textunderscore drn05\textunderscore v03.pdf}}, assuming the T magnitude to be similar to the $I_c$ magnitude, derived earlier. This RMS was multiplied by the square root of two to get the RMS in 30\,min.
Then, we assumed for each target an Earth-sized planet on a close-in orbit (receiving an irradiation of four times that of Earth), and computed the detection SNR expected from two 27\,d observations with TESS.
For the noise of the TESS observations, we took into account the mean number of observed transits during the observation, their depths, durations (assuming a central transit), as well as our estimate for the TESS photometric precision for the star. Furthermore, we assumed a floor noise of 50\,ppm per transit, and no noise for the phase-folded photometry. If the computed SNR was $\ge$5, a detection within the reach of TESS was inferred.

\begin{figure*}[h!]
    \centering
    \includegraphics[width =\columnwidth]{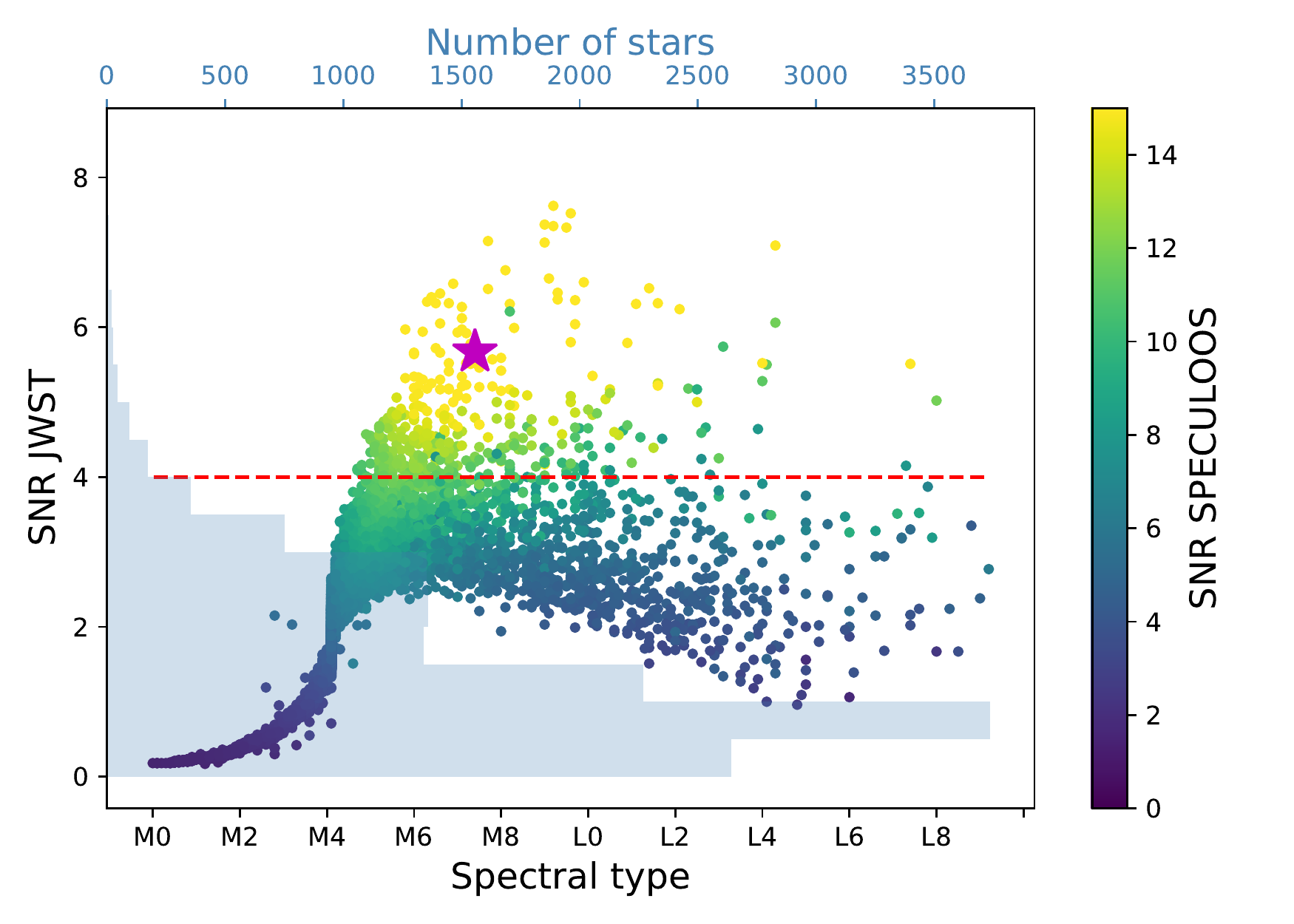}
    \includegraphics[width =\columnwidth]{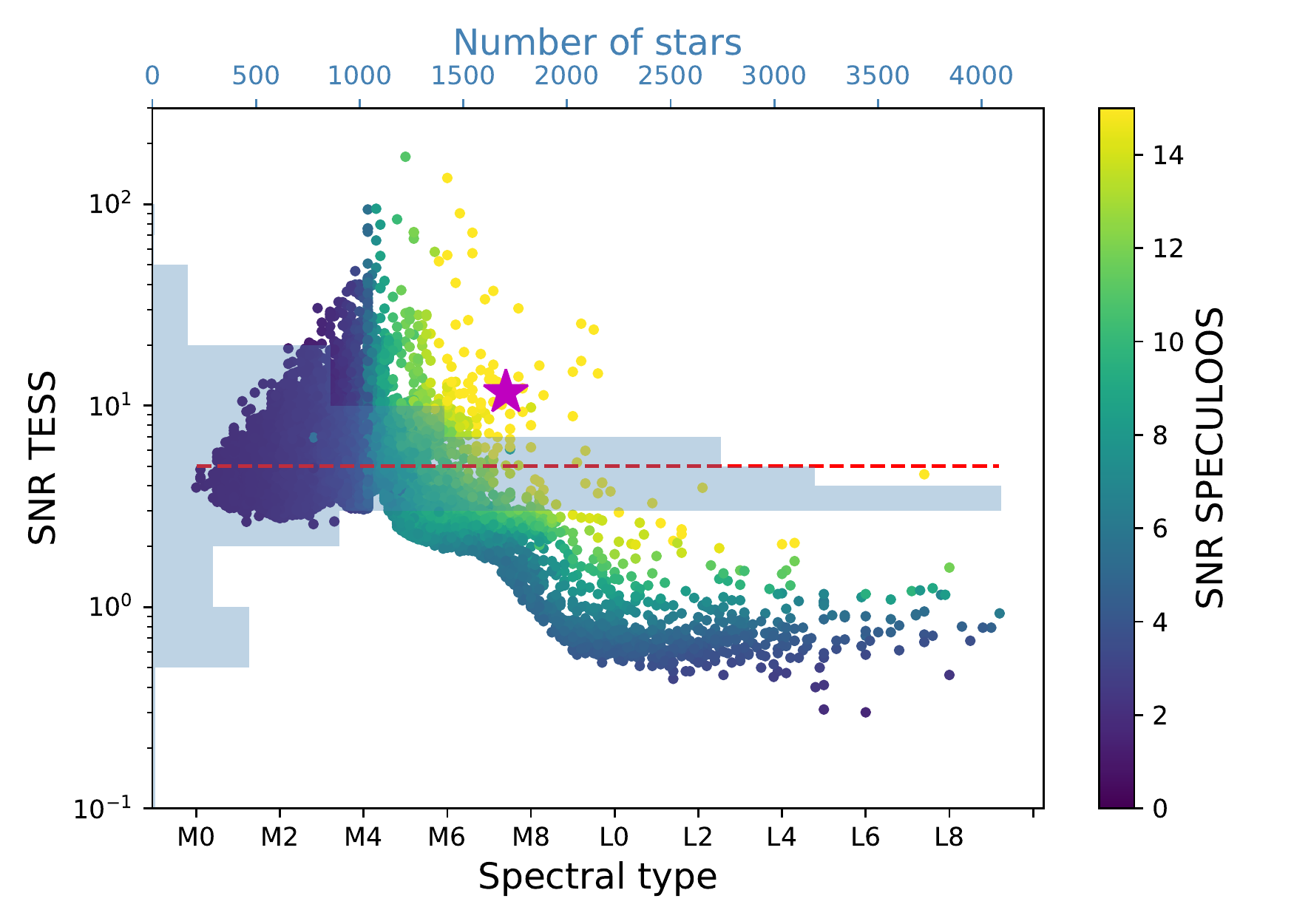}
    \caption{SPECULOOS program selection: {\it Left}: Estimated signal-to-noise ratio (SNR) in transmission spectroscopy with JWST/NIRSPEC (assuming an `Earth-like' planet, 200\,hr of JWST/NIRSPEC time, and a spectral sampling of 100\,nm) as a function of spectral type for the 14,164 M- and L-dwarfs within 40\,pc. TRAPPIST-1 is shown as a magenta star. Red dashed line: SNR = 4, used to select Program~1 targets. 
    Colour coding: estimated SNR to detect a single transit of an Earth-sized planet with one telescope of the SPECULOOS network, with yellow points representing a ${\rm SNR} \geq 15$. Light-blue histogram: Distribution of all objects and their corresponding SNRs.\\ 
    {\it Right}: Estimated signal-to-noise for TESS to detect an Earth-sized planet with an irradiation four times larger than Earth.
    Colour coding: estimated SNR to detect the transit of an Earth-sized planet with the SPECULOOS network, with yellow points representing a ${\rm SNR} \geq 15$. Red dashed line: SNR = 5, used to select Program~2 targets. Light-blue histogram: Distribution of all objects and their corresponding SNR.}
    \label{fig:jwst}
\end{figure*}
We applied the same procedure for SPECULOOS. First, we used the SPECULOOS Exposure Time Calculator to compute the typical photometric precision for each target that should be achieved by a SPECULOOS 1m-telescope in the I+z filter \citep{Delrez2018a} as a function of the apparent $J$-band magnitude and the spectral type of the target. A detailed discussion on the typical noise achieved for SPECULOOS targets with our 1\,m telescopes is given in \cite{murray20}. To simulate this typical noise by using the SPECULOOS Exposure Time Calculator, we assumed a 100\,hr-long photometric monitoring and an average floor noise of 500\,ppm per transit.
Here too, a detection was deemed possible when the computed SNR was $\ge$5.

Figure \ref{fig:jwst} (right) shows a summary of our results. For our 40\,pc sample the detection SNR with TESS for temperate, rocky planets increases for earlier M-dwarfs due to its excellent time coverage and the decreasing orbital periods for temperate planets with increasing spectral type. Nevertheless, it greatly decreases at spectral type $\sim \rm M5$ due to the faintness of the objects, which is in line with \citet{Sullivan2015}. The colour coding shows the SNR, achieved by SPECULOOS for the same temperate planets. The SNR starts to increase at spectral type $\sim \rm M5$ towards later spectral types, due to its shorter monitoring strategy (for most earlier-type objects only one transit can be observed in 100\,hr), but its smaller photon noise compared to TESS.

In this context, for the second program (hereafter Program~2), we selected stars with photometric spectral types M5 and later, for which Earth-sized planets with an irradiation four times that of Earth can be detected by TESS with a SNR $\ge$5. These criteria are met for some targets previously selected in our Program~1. We thus apply those criteria to select stars from our 40\,pc sample that are not in Program~1. 

The target list of Program~2 finally contains 171 objects that have spectral types mainly between M5V and M6.5V. For these targets we aim to identify low-significance (5 to 8 sigma) transit signals of Earth-sized planets in TESS photometry first, which are then observed with SPECULOOS to confirm or discard their planetary nature (see details in Sec. \ref{section:vetting_TESS}).

\subsection{Program~3: The SPECULOOS Statistical Survey}\label{section:program3}
The third program (hereafter Program~3) is the SPECULOOS statistical survey. It focuses on all objects from our target list not covered by the two first programs, including all UCDs (later than M6) from our 40\,pc sample. Given the uncertainty of our classification, we select all objects with photometric classifications of M6 and later to ensure the inclusion of all UCDs.
This program contains 1,121 targets. 

\begin{table}
\centering
\setlength{\tabcolsep}{1.pt} 
\renewcommand{\arraystretch}{1.3} 
\caption{Column description of the SPECULOOS target list.  The complete tables are available for download in the online archive. In all tables the photometric spectral type derived in this work is given as floating point number, starting at 0 for M0. For example 6.5 and 12.0 denote spectral type M6.5 and L2, respectively.}
\label{tab:col_descr}
\footnotesize
\begin{tabular}{lll}
Column name & Unit & Description\\\hline
spc         &                       & SPECULOOS ID\\
twomass     &                       & 2MASS designation\\
gaia        &                       & Gaia DR2 source\_id\\
Simbad\_main\_id &                  & Main identifier for the object\\
Ra          & $\mathrm{deg}$        & Right ascension J2000\\
Dec         & $\mathrm{deg}$        & Declination J2000\\
dist        & $\mathrm{pc}$         & Distance\\
dist\_err   & $\mathrm{pc}$         & Distance error\\
G           & $\mathrm{mag}$        & Gaia DR2 G magnitude\\
G\_err      & $\mathrm{mag}$        & Gaia DR2 G magnitude error\\
I           & $\mathrm{mag}$        & Ic magnitude\\
I\_err      & $\mathrm{mag}$        & Ic magnitude error\\
J           & $\mathrm{mag}$        & 2MASS J magnitude\\
J\_err      & $\mathrm{mag}$        & 2MASS J magnitude Error\\
H           & $\mathrm{mag}$        & 2MASS H magnitude\\
H\_err      & $\mathrm{mag}$        & 2MASS H magnitude Error\\
K           & $\mathrm{mag}$        & 2MASS Ks magnitude\\
K\_err      & $\mathrm{mag}$        & 2MASS Ks magnitude Error\\
bp\_rp      & $\mathrm{mag}$        & Gaia DR2 Colour index\\
spt         &                       & Photometric spectral type\\
spt\_err    &                       & Photometric spectral type error\\
teff        & $\mathrm{K}$          & Effective temperature\\
teff\_err   & $\mathrm{K}$          & Effective temperature error\\
mass        & $\mathrm{M_{\odot}}$  & Stellar mass\\
mass\_err   & $\mathrm{M_{\odot}}$  & Stellar mass error\\
radius      & $\mathrm{R_{\odot}}$  & Stellar radius\\
radius\_err & $\mathrm{R_{\odot}}$  & Stellar radius error\\
BCj         & $\mathrm{mag}$        & Bolometric correction in J\\
BCj\_err    & $\mathrm{mag}$        & Bolometric correction error in J\\
L\_bol      & $\mathrm{L_{\odot}}$  & Stellar Luminosity\\
L\_bol\_err & $\mathrm{L_{\odot}}$  & Stellar Luminosity error\\
spec\_spt   &                       & spectroscopic Spectral type\\
spec\_spt\_ref &                    & Reference for spectroscopic type\\
phot\_spt   &                       & photometric Spectral type\\
phot\_spt\_ref &                    & Reference for photometric type\\
Known\_Binary &                     & Object is a known binary\\
program     &                       & The SPECULOOS observing program\\
SNR\_JWST   &                       & Expected atmospheric SNR for JWST\\ 
SNR\_TESS   &                       & Expected detection SNR with TESS\\
SNR\_SPC    &                       & Expected detection SNR with SPECULOOS\\

\\\hline
\end{tabular}
\end{table}


\section{Catalogue Characterisation and New Targets}

The full SPECULOOS input catalogue is composed of all targets from the three programs described above. 
To characterise the sample, we cross-matched our catalogue with known objects in the literature. As there is no complete database for all UCDs available so far, we cross-checked with several catalogs and databases. The most comprehensive database so far is SIMBAD \citep{wenger00}. It lists spectroscopic classifications for 44\% of our targets. Further catalogues, which we used for the cross-check were: DwarfAchives.org
\footnote{\url{http://spider.ipac.caltech.edu/staff/davy/ARCHIVE/index.shtml}}, the compilation by J. Gagn\'e\footnote{\url{https://jgagneastro.com}}, the list published by \cite{kiman19} and compiled from the BOSS ultracool dwarf survey, the compilation from \cite{Gagliuffi2019}, as well as the compilations from \cite{smart19}, \cite{reyle18}, and \cite{scholz20} based on Gaia DR2. If spectral types from optical and infrared spectra were given, we adopted the spectral type derived from the optical spectrum first and only adopted the infrared spectral type if no optical spectrum was obtained. If a spectral type or catalogue entry was derived from photometry only, we denoted this entry as a photometric spectral type. If the origin of a spectral classification cannot be verified (for example a reference is missing in SIMBAD), we denoted this entry as a photometric spectral type too. 

We rejected the object 2MASS J10280776-6327128 from our catalog, as it is a known white dwarf \citep{kirkpatrick16}. Another three targets were rejected, as they are classified as subdwarfs, namely 2MASS J02302486+1648262 \citep{cruz02}, 2MASS J14390030+1839385 \citep{gizis97}, and 2MASS J17125121-0507249 \citep{aganze16}. Finally, we ended up with a catalogue of 1,657 SPECULOOS targets.

Table \ref{tab:col_descr} shows the list descriptions for the online table. It contains all SPECULOOS targets, their derived parameters as described in Sec. \ref{section:target_sel}, as well as the spectral types available in literature. We find that 50\% have spectral types in literature, 28.1\% have photometric spectral types only, and 21.9\% (363 targets) have been classified in this work for the first time (shown in Table \ref{tab2}\footnote{
The SPECULOOS target catalogue (SPC\_targets) and Table \ref{tab2} are only available in electronic form at the CDS via anonymous ftp to \url{cdsarc.u-strasbg.fr} (130.79.128.5) or via \url{http://cdsweb.u-strasbg.fr/cgi-bin/qcat?J/A+A}.
}). The majority of them (260 targets) are classified as M6 and later. Including the uncertainties from our photometric classification, we refer to those targets as potential `new' UCDs.

\begin{table}
\caption{Comparison between photometric and spectroscopic classifications. The average differences and standard deviations are given in units of sub-classes. Negative values represent earlier photometric classifications, compared to spectroscopic classifications.}
\label{tab3}
\footnotesize
\centering
\begin{tabular}{lrrr}\hline
Spectral type               & M4V - M7V & M8V - L1  & L2 - T2\\\\\hline
Targets compared            & 263       & 403       & 162\\
Average difference          & -0.2      & -0.5      & -1.0\\
Standard deviation          & 0.7       & 0.9       & 1.3\\
\end{tabular}
\end{table}

\begin{table*}
\caption{Newly classified low-mass stars and brown dwarfs from the SPECULOOS input catalogue. The column description is similar to Table \ref{tab:col_descr}
}
\label{tab2}
\begin{footnotesize}

\centering
\begin{tabular}{lllrrccrr}
\hline
spc & twomass & gaia & Ra & Dec & dist & J & spt & spt err \\
 &  &  & $\mathrm{{}^{\circ}}$ & $\mathrm{{}^{\circ}}$ & $\mathrm{pc}$ & $\mathrm{mag}$ &  &  \\\hline
Sp0020+3305 & 00202922+3305081 & 4929042942932181888 & 5.1234310 & 33.0851189 & 12.3 & 10.28 & 5.0 & 0.9 \\
Sp0030+2244 & 00302476+2244492 & 459215468053012096 & 7.6033929 & 22.7469251 & 34.7 & 14.59 & 10.4 & 1.1 \\
Sp0043+2426 & 00433563+2426120 & 5044297634405581696 & 10.8987515 & 24.4367411 & 34.1 & 13.84 & 8.0 & 1.1 \\
Sp0053-3631 & 00531899-3631102 & 65829792677145344 & 13.3291346 & -36.5194963 & 24.1 & 14.45 & 11.3 & 1.1 \\
Sp0103+4509 & 01030791+4509292 & 277483990023898240 & 15.7838542 & 45.1581209 & 23.5 & 13.69 & 9.9 & 1.1 \\
Sp0120-0741 & 01204916-0741036 & 215350076836032384 & 20.2048761 & -7.6843580 & 27.3 & 12.99 & 7.0 & 1.1 \\
Sp0130-4445 & 01303563-4445411 & 2886691573123995520 & 22.6487001 & -44.7614414 & 34.4 & 14.06 & 8.8 & 1.1 \\
Sp0151+6423 & 01515483+6423084 & 3116021729853781376 & 27.9776496 & 64.3856717 & 21.5 & 13.18 & 9.2 & 1.1 \\
Sp0151-0051 & 01514102-0051564 & 5565156633450986752 & 27.9210273 & -0.8656641 & 34.9 & 15.10 & 11.2 & 1.2 \\
Sp0202+1020 & 02021620+1020136 & 3094447525008511488 & 30.5676289 & 10.3372094 & 9.2 & 9.84 & 5.3 & 0.9 \\
\end{tabular}

\end{footnotesize}
\end{table*}

\begin{figure}
    \centering
    \includegraphics[width=\columnwidth]{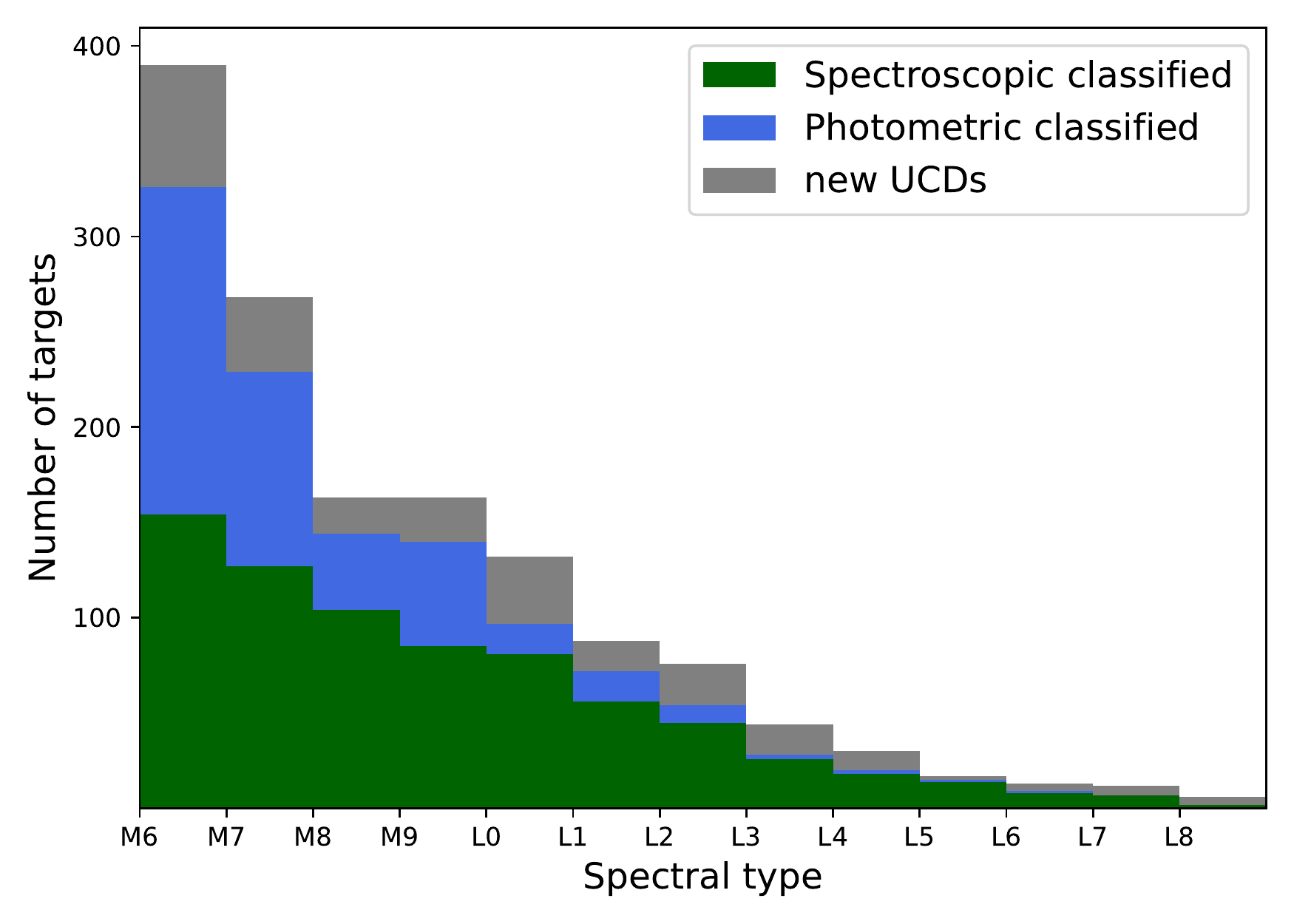}
    \caption{Distribution of photometric spectral types for UCDs within 40\,pc from the SPECULOOS input catalogue. 
    Colour coding: Targets with spectroscopic classification (green), with photometric classification (blue); and newly identified UCDs without spectral type in the literature (gray).}
    \label{fig:SPC_UCDs}
\end{figure}

The spectral type distribution of the SPECULOOS catalogue peaks at M6V with about 400 targets and decreases with later spectral types. As shown in Fig. \ref{fig:cat}, The catalogue is incomplete for targets earlier than M6V because of the varying spectral type cut of the different programs. Nevertheless, no cuts were introduced for later-type stars. In Fig. \ref{fig:SPC_UCDs} we show the distribution of UCDs within catalogue (photometric spectral type M6V or later) and the corresponding coverage with literature spectral types. 
We note an apparent lack of spectral classifications for earlier-type objects, while the sample is almost complete for BDs. This bias is in line with the findings by \cite{reyle18}. It origins mostly from intensive spectral typing of the BD sample in the past decade.

We compared our photometric spectral types (derived as described in Sec. \ref{section:target_sel}) to spectroscopic measurements for the 829 targets for which we found a spectroscopic classification in the literature. 
We found that the photometric classification agrees best for the M-dwarf stars in our sample with an average difference below one sub-class. For later objects and brown dwarfs, we found that the photometric  classification is on average one sub-class earlier than the spectroscopic classifications. The results of the comparison are shown in Tab. \ref{tab3}.

\section{Observing Strategy}\label{section:strategy}

\subsection{SPECULOOS monitoring strategy}\label{section:strategy1}

The basic observational strategy of SPECULOOS is to observe each target continuously with one telescope for a duration long enough to make likely the observation of at least one transit of a temperate planet.
Therefore, we observe with each telescope one or two targets per night. Other surveys have chosen to proceed differently \citep{Nutzman2008,tamburo19}, but considering our requirement for very-high photometric precision and the expected short transit duration (down to 15\,min) for very-short-period ($\leq$ 1\,d) planets orbiting UCDs, we estimate that a continuous observation approach is more appropriate (\cite{Delrez2018b,gibbs20}).

Considering that we have divided our target list in three distinct programs, SPECULOOS's observing strategy is also divided in three. 
For instance, Program~1 aims to perform a more completed coverage of the habitable zone of its targets than Program~2 and 3, and so its monitoring duration per target has to be larger than for the two other programs. 

Until Nov. 2019, our strategy was the same for all programs, with a monitoring duration of 100\,hr for any target no matter which program it belongs to (that is to say stars were observed until 100\,hr of photometric data have been gathered). Furthermore, we used to split the observation for each target into two observation blocks of each 50\,hr per year. Our intention back then was to survey many targets to detect quickly very-short-period planets, which would enable us to trigger intensive follow-up to detect additional planets (if present) with larger periods in the system. However, we have now settled on a different strategy: a monitoring duration of 200\,hr in one block for Program~1's target, and a monitoring duration of 100\,hr in one block for Program~2 and 3's targets. 

\begin{figure}[h!]
    \centering
    \includegraphics[width=\columnwidth]{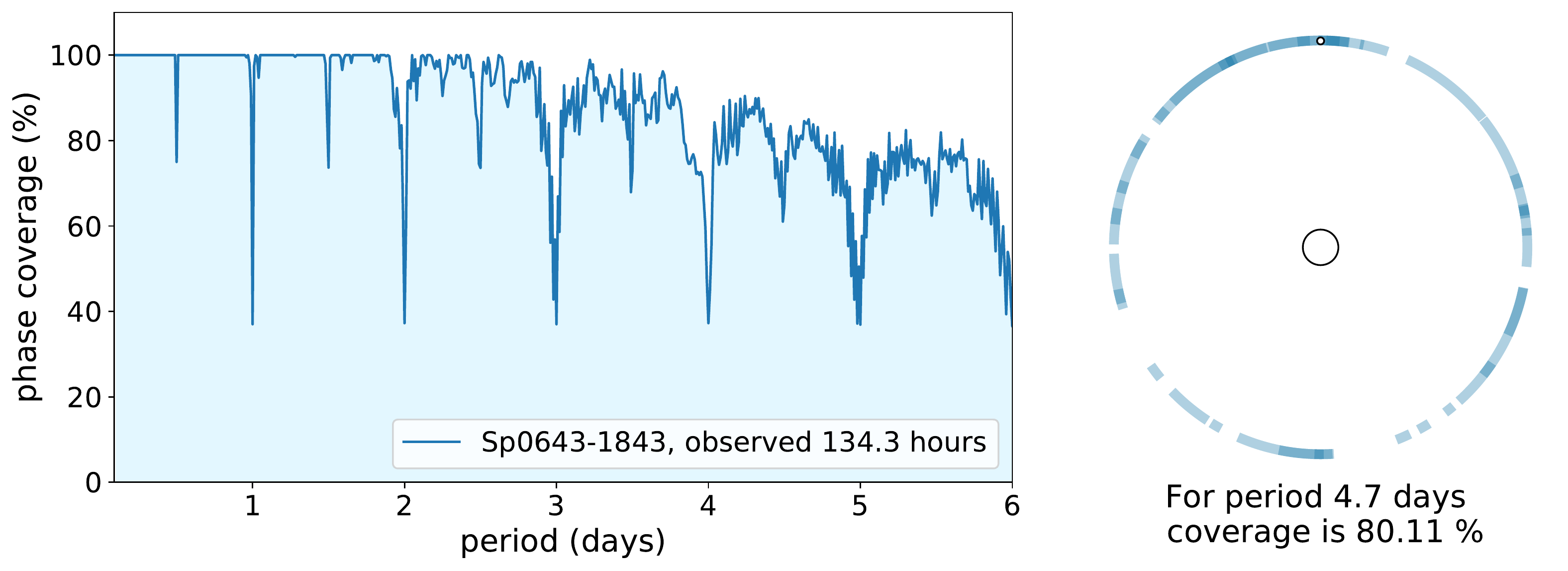}
    \caption{{\it Left:} Phase coverage as a function of the orbital period of an hypothetical planet around the target {\fontfamily{pcr}\selectfont Sp0643-1843} (chosen arbitrarily), observed for 134.3 hours in total by SPECULOOS.
    {\it Right:} Graphical visualisation of the coverage of target {\fontfamily{pcr}\selectfont Sp0643-1843} for an orbital period of 4.7 days. Each blue circular arc represents one night of observation; its size is proportional to the number of hours observed each night and the full circle depicts a duration of 4.7 days. } 
    \label{fig:example_coverage}
\end{figure}

To ensure we picked the most appropriate strategy for the three SPECULOOS observation programs, we used a metric that we call the effective phase coverage. This metric gives an estimation of how the phase of an hypothetical planet would be covered for a range of periods, using existing observations from the SPECULOOS network. More precisely, we computed the percentage of phase covered for each orbital period from $P=0.1$\,d to $P=P_{max}$ and took the effective phase coverage for periods $\leq$ $P_{max}$ to be the integral over the period range. Figure \ref{fig:example_coverage} shows the phase coverage for each possible orbital period for an arbitrary target observed for 134.3 hours by SPECULOOS. The effective phase coverage is depicted by the blue area, for $P_{max}=6$\,d.

In the next sections we detail how this metric drives our program-specific strategy.

\subsubsection{Strategy of Program~2 and ~3} \label{section:strategy_prog2&3}

Programs~2 and 3 focus on the detection of planets with irradiations similar to TRAPPIST-1b (4\,S$_{\oplus}$) and thus with short orbital periods. To ensure that we cover all short-period planets, we set $P_{max}=6$\,d and calculated the effective phase coverage for all targets for which observations with the SPECULOOS network have been started. Additionally we performed simulations of SPECULOOS observations, assuming 4 hours of observations per night per target and losses (bad weather or full moon too close) of 30\%.

\begin{figure}[ht!]
    \centering
    \includegraphics[width=0.8\columnwidth]{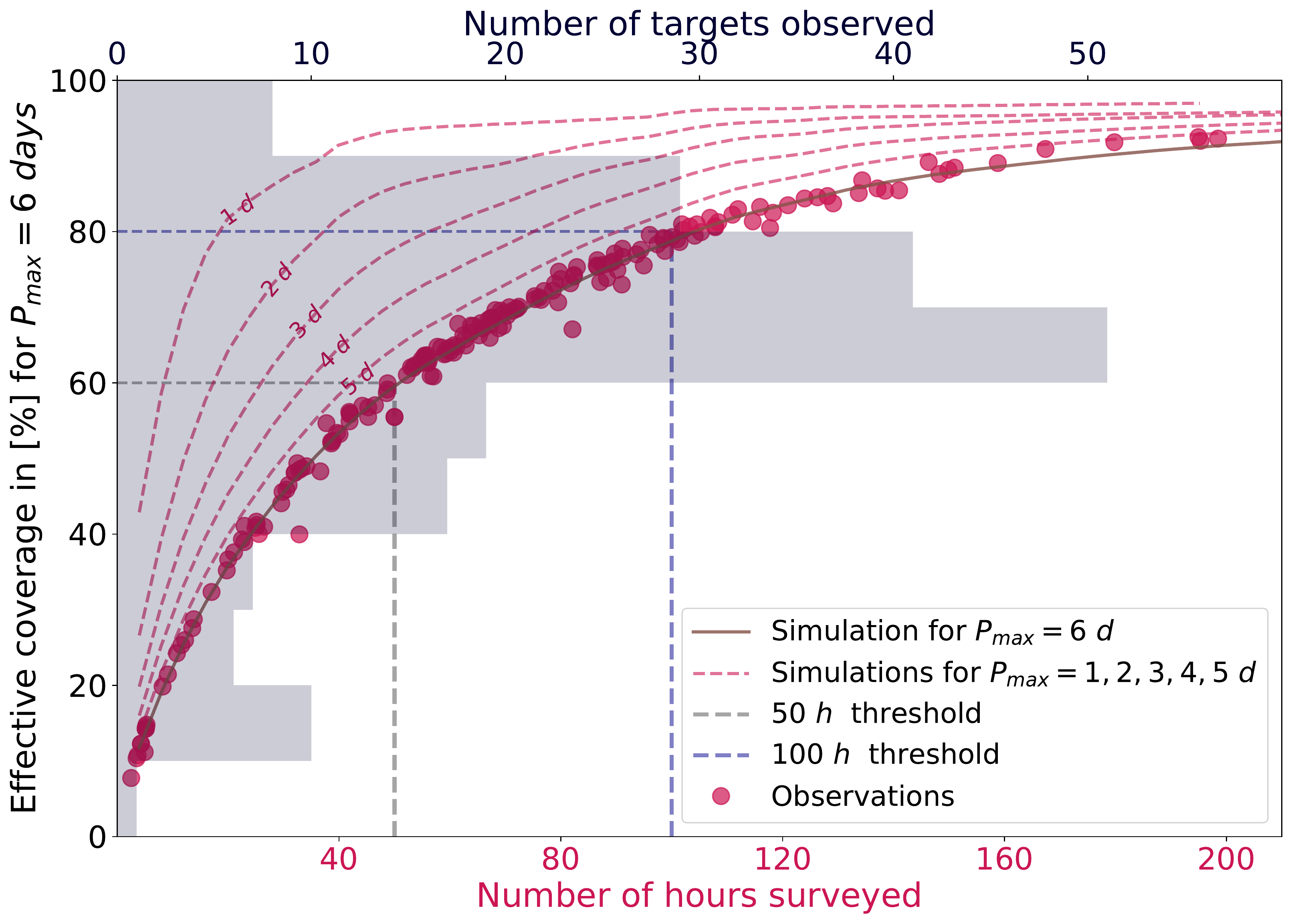}
    \caption{Effective coverage as a function of the number of hours surveyed for all targets observed by the SPECULOOS telescopes so far. Coloured dots shows the coverage calculated from existing observation  for periods going from 0.1 to 6 days, and the solid red line is the corresponding simulation. Dashed red lines are simulations for various values of $P_{max}$ ranging from 1 to 5 days. Gray histogram: Number of targets in each slice of coverage. }
    \label{fig:cov_vs_hoursobs}
\end{figure}

Figure \ref{fig:cov_vs_hoursobs} shows the evolution of the effective coverage as a function of the number of hours surveyed for all periods $\leq$ $P_{max}=6$ days.
We observe that for 50 hours of observations the expected phase coverage is ~60\%, whereas for 100 hours of observations it increases to ~80\%. 
The SPECULOOS data agree well with our simulations and show that for our targets in Programs~2 and ~3 a monitoring duration of 100\,hr will allow us to detect close-in planets with an effective phase coverage of more than 80\%. 
Figure \ref{fig:cov_vs_hoursobs} also shows the number of targets for which observations have started as a function of the coverage. We see that the majority of these targets have been covered at 60-70\% for periods going from 0.1 to 6 days. This is explained by the fact that for many targets we already gathered more than 50 hours during the first year of observations.

\begin{figure}[h!]
    \centering
    \includegraphics[width=0.8\columnwidth]{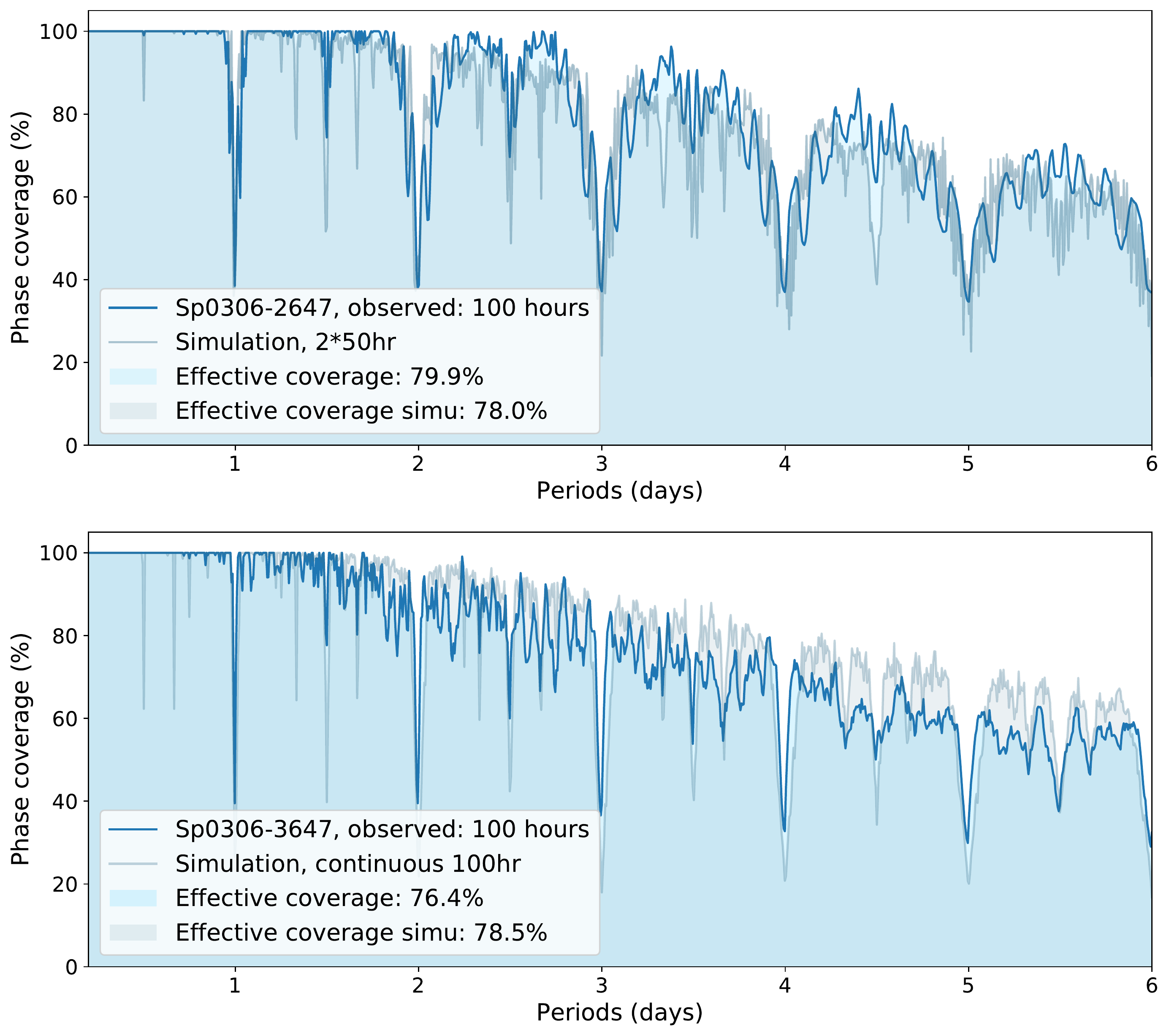}
    \caption{{\it Top:} Evolution of the phase coverage as a function of the period for one SPECULOOS target ({\fontfamily{pcr}\selectfont Sp0306-2647}) observed for 100\,hr with SSO in two blocks of 50hr several months apart. The solid blue line shows the evolution of the coverage calculated from existing observations whereas the gray line is the result of simulations.
    {\it Bottom:} Same for another SPECULOOS target ({\fontfamily{pcr}\selectfont Sp0306-3647}) observed 100\,hr with SSO but on consecutive days. } 
    \label{fig:comparison_strategies}
\end{figure}

To analyse the effect of splitting the 100\,hr observations into two blocks for 50\,hr per year, we compared the effective phase coverage for two SPECULOOS targets that have both been monitored with SPECULOOS for 100\,hr, but with the two different strategies. Figure \ref{fig:comparison_strategies} shows the phase coverage for each possible orbital period for both targets. Also shown are simulated SPECULOOS observations, for comparison.

In the first scenario, the target was observed for 50\,hr on consecutive days, and again 50\,hr one year later (top panel of Figure \ref{fig:comparison_strategies}), whereas in the second scenario the target was observed day after day until 100hr of observations were reached (bottom panel of Figure \ref{fig:comparison_strategies}).
We found that splitting the observations better shuffles the phases but also introduces aliases that result in a slightly decreased coverage for orbital periods close to multiples of sidereal days. As a whole, observations and simulations indicate that the effective coverage is with $-0.7 \pm 1.4\%$ similar for both scenarios.

More generally, in our simulations we used a period grid of 15\,min which resembles the minimum expected transit duration. We find that between 0.1 and 6 days, about 40\% of all possible orbital periods gain phase coverage from continuous observations, compared to 20.0\% that loose coverage. Despite similar effective phase coverage, for about 20\% (30\% for planets with 8-12\,d period) of all possible orbital periods we find a better phase coverage for continuous observations with SPECULOOS.

Besides, as the continuous observations strategy is also more straight forward from the scheduling point of view, we opted to observe each target without additional splitting until its program-specific monitoring duration is reached.
In the same way, we expect a slight increase of effective phase coverage of $1.2 \pm 0.7\%$ for observations in the case that we can observe a target all night long. We also find an increased phase coverage for about 7\% of all possible orbital periods. We thus, opted to observe each target the whole night, if possible.

\subsubsection{Strategy of Program~1}

The objective of our Program~1 is to detect putative transiting `Earth-like' planets with irradiations similar to Earth (1\,S$_{\oplus}$) and thus with orbital periods in or close to the habitable zone (HZ) of its host.

\begin{figure}
    \centering
    \includegraphics[width=\columnwidth]{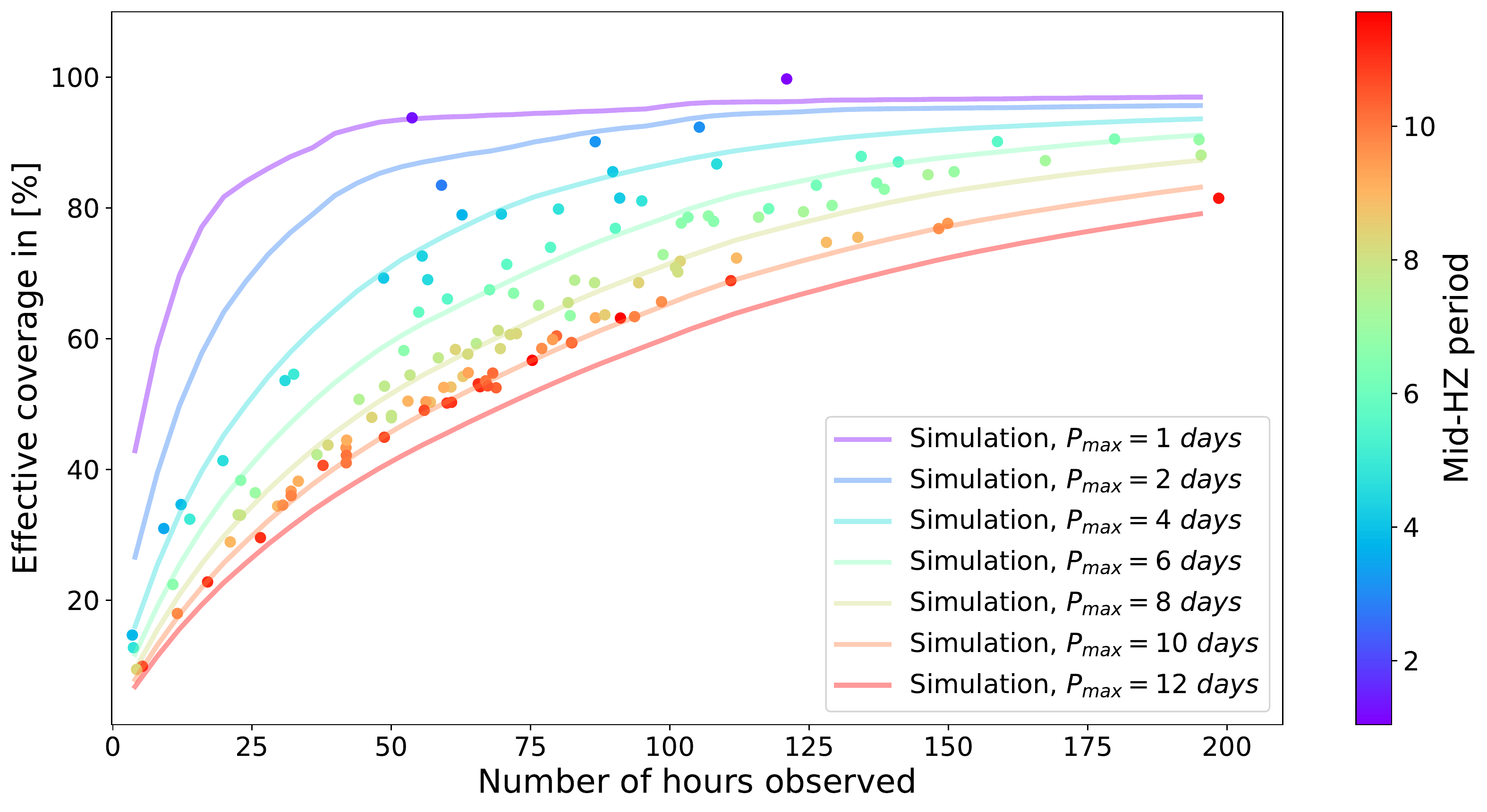}
    \caption{Coverage of the middle of the HZ for Program~1's targets observed so far by the SPECULOOS survey. Solid lines: theoretical simulations, Coloured dots: actual SPECULOOS observations. Colour coding: The expected maximum orbital periods associated with the middle of the HZ of each target.}
    \label{fig:cov_HZ}
\end{figure}

Figure \ref{fig:cov_HZ} shows the effective coverage of Program~1's targets already started with SPECULOOS as a function of the number of hours surveyed. In here, we use as $P_{max}$ the middle of the HZ), which has been calculated for each target following \cite{kopparapu13}.
For most of the observed Program~1's targets a hypothetical planet in the HZ would have a period of 8 to 10 days and would require $\simeq$ 200\,hr of observation to reach a coverage $\geq$ 80\%. Thus, for our targets in Program~1 a monitoring duration of 200\,hr will effectively allow us to detect planets in the habitable zone. Besides, the conclusion given in Section \ref{section:strategy_prog2&3} and still true for 200\,hr, such that we decided to monitor each target 
on consecutive days if possible, until the monitoring duration has been reached, without splitting the observations in different years.


\subsection{SPECULOOS targets observed with TESS} \label{section:vetting_TESS}

We selected in Program~2 all objects that would allow the detection of temperate (4 times the irradiation of the Earth), Earth-sized planets with TESS. This is also the case for many targets in our Program~1, as for the earliest-type of them, TESS photometry should be precise enough to enable the detection of short-period planets. For the others, it could at least constrain their photometric variability. We thus decided to analyse systematically the TESS photometry in advance (if possible) of our SPECULOOS observations, and eventually to perform a global analysis of the light curves gathered by both surveys so to maximize the scientific return.

At the time of writing (May 2020), from the targets in our Programs~1 and 2, 231 have 2-min cadence lightcurves provided by SPOC (Science Processing Operations Center, \cite{jenkins16}), 167 have only 30-min cadence lightcurves from Full Frame Images, and 138 are scheduled to be observed by TESS. None of the targets observed have been issued an alert from the automatic pipelines of TESS. This hints to a lack of clear transits in TESS's data set.

In order to search for threshold-crossing events which might be unnoticed, we make use of our custom pipeline {\fontfamily{pcr}\selectfont  SHERLOCK} \footnote{{\fontfamily{pcr}\selectfont  SHERLOCK} (\textbf{S}earching for \textbf{H}ints of \textbf{E}xoplanets f\textbf{R}om \textbf{L}ightcurves \textbf{O}f spa\textbf{C}e-based see\textbf{K}ers)
code is available upon request.} \citep{pozuelos2020,demory2020}. This pipeline downloads the Pre-search Data Conditioning Simple Aperture (PDC-SAP) flux data from the NASA Mikulski Archive for Space Telescope (MAST) by means of the {\sc{lightkurve}} package \citep{lightkurve2018}. Then, it removes outliers defined as data points $>3\sigma$ above the running mean, and removes the stellar variability using {\sc{wotan}} \citep{Hippke2019Python} using two different detrending methods: a bi-weight filter and a Gaussian process with a Matern 3/2-kernel. To optimize the search for transits, the pipeline is run for a number of trials in which the window and kernel sizes are varied for each respective aforementioned method.
The transit search is performed by the {\sc{transit least squares}} package \citep{Hippke2019TransitPlanets}, which is optimized for the detection of shallow periodic transits. We set our threshold limits at SNRs of $\geq5$ and signal detection efficiencies (SDEs) of $\geq5$, and follow the in-depth vetting process described by \cite{Heller2019TransitK2}.

For each event that overcomes the vetting process, we perform a ground-based follow-up campaign with our SPECULOOS network to rule out potential sources of false positives and strengthen the evidence of its planetary nature. Given the large pixel size of TESS (21``~pixel$^{-1}$), the point spread function can be as large as 1', which increases the probability of contamination due to a nearby eclipsing binary (EB). Indeed, a deep eclipse in a nearby EB may mimic a shallow transit detected for a target star due to dilution. Hence, it is critical to identify potential contamination due to EBs and/or relatively nearby neighbours up to 2.5~arcmin \cite[e.g.][]{kostov2019,quinn2019,gunther2019,nowak2020}. To schedule photometric time-series follow-up observations, we use the TESS Transit Finder, which is a customized version of the Tapir software package \citep{jensen2013}, in coordination with the SPECULOOS scheduler described in Section \ref{section:scheduler}. 

\begin{figure}
    \centering
    \includegraphics[width =\columnwidth]{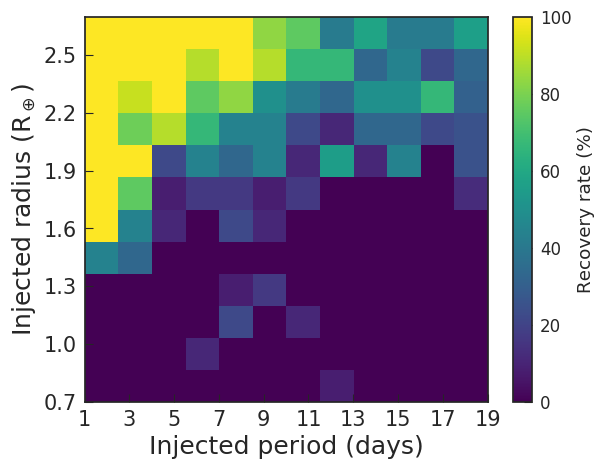}
    \caption{Graphical example of the inject-an-recovery test performed for one of our candidates in Program~1 (Sp0025+5422). The R$_{\mathrm{planet}}$--P$_{\mathrm{planet}}$ parameter space explored corresponds to a planetary radius range of 0.7--2.7~R$_{\oplus}$ with steps of 0.05~R$_{\oplus}$, and a period range of 1--19~d with steps of 0.5~d. This yields a total of 1,517 scenarios.
    The recovery rate is presented by the colour coding. Planets smaller than 1.4~R$_{\oplus}$ would remain undetected for almost the full set of periods explored. 
    }
    \label{fig:recovery}
\end{figure}

For the analysis of a given candidate that yields negative results, we may wonder if the photometric precision of the TESS lightcurve is high enough to confidently rule out the presence of Earth-sized planets in the habitable zone. To check the detectability of such planets in the TESS data set, we performed an injection-and-recovery test. This consists of injecting synthetic planetary signals into the PDC-SAP flux light curve before the detrending and the transit search stages described above. We explore the R$_{\mathrm{planet}}$--P$_{\mathrm{planet}}$ parameter space in the ranges of 0.7--2.7~R$_{\oplus}$ with steps of 0.05~R$_{\oplus}$, and 1--$x$~d with steps of 0.5~d, where $x$ is the largest period for a planet residing in the habitable zone of the given candidate \citep{kopparapu13}. For example, for the TRAPPIST-1 system, $x\sim$15~d, which means an exploration of periods ranging from 1 to 15~d, which combined with radius range (0.7--2.7~R$_{\oplus}$) consists of 1,189 different scenarios. A graphical example is shown in Fig.~\ref{fig:recovery} for one of our candidates in Program~1, which yielded a negative result in the previous search for threshold-crossing events.
If after all these analyses, we cannot firmly rule out the presence of Earth-sized planets orbiting in the habitable zone of a given star in our list, we will observe it for 100\,hr with the SPECULOOS network. 

\section{SPECULOOS scheduler}\label{section:scheduler}

In order to handle the three different programs and optimise the visibility windows of our 1,657 targets, we have developed a planification tool, named SPeculoos Observatory sChedule maKer ({\fontfamily{pcr}\selectfont{SPOCK})}. All SPECULOOS observatories are controlled using the commercial software ACP\footnote{\url{http://acp.dc3.com}}, thus its first aim is to prepare daily observing scripts (ACP plans) for the SSO, SNO and SAINT-EX observatories (source code available on Github\footnote{\url{https://github.com/educrot/SPOCK}}).  

To decide which targets to observe, {\fontfamily{pcr}\selectfont  SPOCK} relies on several criteria:\\
- priority; essentially the value of the SNR for JWST transmission spectroscopy for an `Earth-like' planet in Program~1, the TESS detection SNR for a temperate planet in Program~2, and the SPECULOOS detection SNR for a temperate planet in Program~3.\\
- observability; SPOCK computes the best visibility window of the year for each target. Every time a schedule is made, SPOCK selects new targets that are at their optimum visibility at this time of the year. The selected targets are then ranked and the one with the highest priority is scheduled (providing it respects constraints imposed by the facility like moon distance and altitude). If observable all night, the target is simply scheduled all night, but if some gaps remain an additional target is added to complement the schedule and avoid losing observing time. Furthermore, to prevent having too short observation blocks (1\,h or less) the duration of the two targets are set to be comparable. For instance, if one night is 8\,h long and target~1 is observable for the first 7\,h only, target~2 is not going to be scheduled for the last hour only but rather the night will be split in half such that target 1 and 2 are observed for approximately the same amount of time. We say approximately because we do not exactly split the night in half, instead we adopt nightly observing duration to each target's visibility which will shift from night to night during the observation period.
This situation of observing two targets per night rather than one can happen frequently as many targets have latitudes that do not allow to fill up all the night time available for the given site, and this even at their peak of visibility. As a whole, we can make the assumption that a night lasts 8\,h, and approximate that the duration of the observation for each target is 4\,h, which justifies why we used observation blocks of 4\,h in the simulations we carried out in Section \ref{section:strategy1} 

To implement those constraints, {\fontfamily{pcr}\selectfont  SPOCK} makes use of the {\fontfamily{pcr}\selectfont ASTROPLAN} package \citep{astroplan2018}, a flexible Python 
toolbox for astronomical observation planning and scheduling. {\fontfamily{pcr}\selectfont  SPOCK} also optimizes on the period of the year for which the target is the most visible at a relatively low airmass.\\
- completion ratio $r_{c}=\frac{hours_{observed}}{hours_{threshold}}$; this ratio embodies the fraction of hours of observation completed versus the number of hours required for each target. Note that the value of $hours_{threshold}$ depends on the program to which the target belongs, 200 hr for Program~1 and 100\,hr for Program~2 and 3. Using this completion ratio to rank targets is useful to favour the quick completion of on-going targets rather than starting new ones continually.\\
- coordination; as SPECULOOS is an all-sky survey, one of the main roles of {\fontfamily{pcr}\selectfont  SPOCK} is to handle the coordination of multi-site observations. For instance, between two targets with similar priority but one observable only from one site and the other from several sites, {\fontfamily{pcr}\selectfont  SPOCK} will choose the target that yields the most coverage. Besides, when possible, 1 hour overlap between observations from two different sites is scheduled to help the recombination of the light curves. 

\begin{figure}[h!]
    \centering
    \includegraphics[width=0.9\columnwidth]{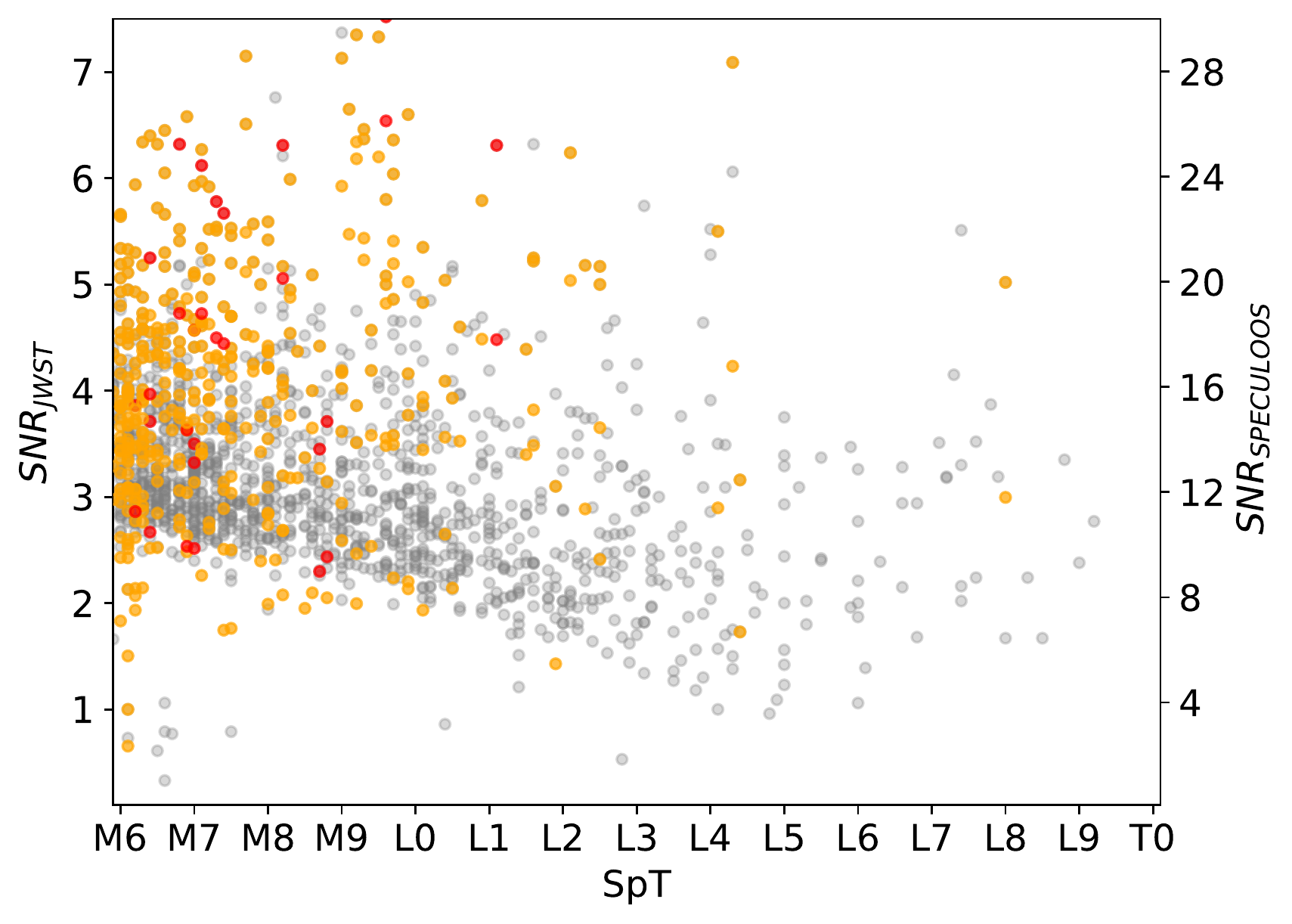}
    \caption{Estimated signal-to-noise ratio (SNR) as a function of spectral type. Gray dots represents all targets in the SPECULOOS target list, orange dots stand for targets for which observations have been initiated but not completed, and red dots stand for completed targets (i.e., observed for 200\,hr if the target is in Program~1 and for 100\,hr if the target is in Program~2 or 3).}
    \label{fig:spock}
\end{figure}

Figure \ref{fig:spock} shows the targets scheduled so far by {\fontfamily{pcr}\selectfont  SPOCK}. As explained in Section \ref{section:strategy}, the former strategy of dividing the observations in blocks of 50hr was prevailing until recently (Nov 2019). Therefore many targets are currently in the "on going observation" phase. We note that observations have started for more than half of Program~1 targets ($\rm SNR_{JWST}$ $\geq$ 4).

\section{Updated planet-yield for SPECULOOS}

In \cite{Delrez2018b} (hereafter D18), we simulated the planet yield expected from the SPECULOOS survey, assuming a total observation of 100\,hr per target. We updated this simulation in order to include the SPECULOOS target list presented in this paper. Also, we accounted for the physical parameters, derived for each target.

To estimate the planet detection yield, we used a Monte Carlo simulation that looped 1,000 times over the target sample. First, we assumed that every target has at least one planet and used the mass, radius, and effective temperature as starting points to simulate a planetary system. To account for the errors in stellar parameters, in each loop, we drew the simulated mass, radius, and effective temperature from a normal distribution with $\mu= \text{parameter}$ and $\sigma = \text{error of parameter}$. 

Second, we draw for each target a planetary system after \cite{miguel20}, by using their distribution of expected number of planets. We accounted for the period distribution by using two period samples, one between 0.5 and 23 days and a second between 23 and 1800 days. Drawn planets are uniformly distributed with a probability of p=0.5 to be in one of both samples. The planet masses and densities are drawn from a normal distribution based on the TRAPPIST-1 system with:
\begin{itemize}
    \item Planet Mass $(M_{\oplus})$  N(0.81 , $0.34^2$)
    \item Planet density $(\rho_{\oplus})$ N(0.79 , $0.12^2$) 
\end{itemize}

We derived the planet parameters and inclinations based on the stellar parameters using the same methods as described in D18. Simulated planets are counted as 'discovered', if following criteria are met: (1) A transit has been observed during a simulated SPECULOOS observation run. This assumes a visibility of 4\,hr per night, a total loss of about 30\% due to weather or technical downtime, and a monitoring campaign that completes after the program-specific number of observing hours has been reached. (2) The SNR of the transit is larger than 5, assuming a floor noise during the transit of 500\,ppm. 
From this simulation, we derived: (i) the number of planets detected during the SPECULOOS observation windows; (ii) how many of them are in fact transiting systems (having more than one transiting planet); and (iii) how many of them are in the habitable zone. We denote a simulated planet as being in the habitable zone if the received incident flux meets the criterion: $0.2 < S_{\oplus} < 0.8$, which resembles the conservative limits derived by \cite{kopparapu13}.

Table \ref{tab:planet_yield} summarises the results for the three SPECULOOS sub-programs, as well as the results from D18 for comparison. For the complete SPECULOOS program, with all observing programs finished, we expect $29\pm4$ planets, with $8\pm2$ of them being in the habitable zone of their host star. 
The changes to our expectation from D18 are mainly due to the different simulated system architectures. In D18, we assumed about the same number of planets, but distributed within a period range extending only to 23\,d. Planets with large periods are less likely to be transiting. For the D18 expectation about 1.8\% of all planets were transiting, while we now find a fraction of 0.9\% of transiting planets. Furthermore, SPECULOOS is by design less sensitive to planets with long periods.
We find that 6.5\% of all targets with at least one transiting planet host seven and more planets ($0.5\%$ for $Kmag < 10.5$). Given the expected number of detections, we conclude that the detection by SPECULOOS of a `better' system than TRAPPIST-1 (in terms of atmospheric follow-up potential) is very unlikely.

\begin{table}
	\centering
	\footnotesize
	\caption{Expected planet yield from the SPECULOOS survey.}
	\label{tab:planet_yield}
	\begin{tabular}{lcccc} 
		\hline
		Sample          & Program~1 & Program~2 & Program~3 & D18\\
		\hline
		Targets         & 365       & 171       & 1121      & 1136    \\
		Hours obs.      & 200       & 100       & 100       & 100    \\
		Planets         & $10\pm5$  & $4\pm3$   & $15\pm5$  & $42\pm10$ \\
		Systems         & $3\pm2$   & $2\pm1$   & $7\pm2$   & $22\pm5$ \\
		HZ planets   & $3\pm3$   & $1\pm1$   & $4\pm3$   & $14\pm5$ \\
		\hline
	\end{tabular}
\end{table}

At the time of writing (May, 2020), we monitored about 10\% of all SPECULOOS targets for more than 50\,hr. Using our simulation with this sub-sample, we would expect about $2 \pm 2$ planets to be discovered in the survey by now, which is consistent to the sole detection of TRAPPIST-1 so far.

\section{Summary and Conclusions}

We have presented here the SPECULOOS target list that contains 1,657 very-low-mass dwarfs within 40\,pc. We have derived photometric classifications and stellar properties for each target. 

Our target selection procedure is based on a cross-match between Gaia DR2 and 2MASS catalogues, not only in terms of astrometric position but also in terms of inferred spectral type. We show that our sample is indeed volume limited for UCDs with spectral type earlier than M8, which includes the brightest UCDs and thus the most promising targets to search for terrestrial planets, suitable for atmospheric characterisation.
The rationale of our target selection is to minimize the false positive rate, which is important, as about 50\% of our targets lack a spectroscopic classification and only a few of the spectroscopically classified targets have been identified as false positives like white dwarfs or subdwarfs. 
The SPECULOOS target list includes targets with photometric spectral types from M4 to L9. 363 targets have not been classified in the literature so far, with 260 of them having photometric spectral types of M6V and later. Including the uncertainties of our classification, we denote those targets as potential `new' UCDs.

The main goals of the SPECULOOS project are (1) to detect `Earth-like' planets that are perfectly suited for atmospheric characterisation with JWST and (2) to draw constraints to the structure of planetary systems of UCDs, by surveying our volume limited sample of UCDs for exoplanet transits.

To optimise the scientific return of the project, we divided it into three, non-overlapping observation programs: Program~1, which includes all targets that allow transit transmission spectroscopy with JWST for `Earth-like' (irradiation of 1\,$S_{\oplus}$) planets (365 targets); Program~2, which includes all targets that allow a detection of Earth-sized temperate (irradiation of 4\,$S_{\oplus}$) planets -- like TRAPPIST-1b -- with TESS (171 targets); and Program~3, which includes all other SPECULOOS targets with spectral type M6 and later, which aims to explore the planet occurrence rate for ultracool dwarfs within our 40\,pc sample (1,121 targets).
From all targets in Program~1, we expect for only 44 of them a higher SNR than for the TRAPPIST-1 planets. This underlines the unique nature of this system and its importance for atmospheric characterisation of `Earth-like' planets.

Our strategy is to observe all targets, reaching an effective phase coverage of up to 80\% for temperate planets in Programs~2 and~3 and for planets in the habitable zone in Program~1, resulting in monitoring durations with our SPECULOOS telescope network of 100 to 200\,hr respectively. To optimise the survey, we leverage on the synergy with TESS by analysing TESS data for all targets in our Programs~1 and~2 that (1) have been observed by the TESS mission, and (2) are bright enough that we can discover temperate Earth-sized planets in TESS.

Furthermore, we derived the expected planet yield from the SPECULOOS survey, using a Monte Carlo simulation based on our target list and observation programs. Given our observational strategy, we expect SPECULOOS to detect a few dozen planets, including up to a dozen potentially habitable ones. We also find that with SPECULOOS we should have already found $2 \pm 2$ planets, among our targets. Within the errors, this is consistent with our current lack of detections, apart from TRAPPIST-1. 

Given its focus on Earth-sized planets in a volume limited all-sky sample of UCDs, the detection rate inferred from the SPECULOOS survey will allow us to draw robust constraints to the structure of planet systems of UCDs.

As 50\% of our targets were not spectroscopically classified in literature yet, a large scale spectroscopic follow-up program will be necessary in order to (1) to improve our photometric spectral classification and (2) to exclude potential false classifications.  
The upcoming Gaia DR3 release will largely improve the parallaxes and colours for the 40\,pc sample and, thus, include more targets that were excluded in this publication. We plan to use our target selection algorithm to identify further nearby targets once the next release will be available.

\begin{acknowledgements}

The research leading to these results has received funding from the European Research Council (ERC) under the European Union's Seventh Framework Programme (FP/2007--2013) ERC Grant Agreement n$^{\circ}$ 336480, from the ARC grant for Concerted Research Actions financed by the Wallonia-Brussels Federation, from the Balzan Prize Foundation, and from F.R.S-FNRS (Research Project ID T010920F). MG is F.R.S.-FNRS Senior Research Associate.
This research is also supported work funded from the European Research Council (ERC) the European Union’s Horizon 2020 research and innovation programme (grant agreement n$^{\circ}$ 803193/BEBOP), from the MERAC foundation, and through STFC grants n${^\circ}$ ST/S00193X/1 and ST/S00305/1. 
B.V.R. thanks the Heising-Simons Foundation for Support.
This work has made use of data from the European Space Agency (ESA) mission
{\it Gaia} (\url{https://www.cosmos.esa.int/gaia}), processed by the {\it Gaia}
Data Processing and Analysis Consortium (DPAC,
\url{https://www.cosmos.esa.int/web/gaia/dpac/consortium}). Funding for the DPAC
has been provided by national institutions, in particular the institutions
participating in the {\it Gaia} Multilateral Agreement.
This publication makes use of data products from the Two Micron All Sky Survey,
which is a joint project of the University of Massachusetts and the Infrared Processing 
and Analysis Center/California Institute of Technology, funded by the National Aeronautics 
and Space Administration and the National Science Foundation.
This research has made use of the SIMBAD database,
operated at CDS, Strasbourg, France, This research has made use of the VizieR catalogue access tool, CDS, Strasbourg, France, This research made use of Astropy,\footnote{http://www.astropy.org} a community-developed core Python package for Astronomy \citep{astropy:2013, astropy:2018}.

\end{acknowledgements}


\begin{thebibliography}
\expandafter\ifx\csname natexlab\endcsname\relax\def\natexlab#1{#1}\fi

\bibitem[{{Aganze} {et~al.}(2016){Aganze}, {Burgasser}, {Faherty}, {Choban},
  {Escala}, {Lopez}, {Jin}, {Tamiya}, {Tallis}, \& {Rockward}}]{aganze16}
{Aganze}, C., {Burgasser}, A.~J., {Faherty}, J.~K., {et~al.} 2016, \aj, 151, 46

\bibitem[{{Alibert} \& {Benz}(2017)}]{alibert17}
{Alibert}, Y. \& {Benz}, W. 2017, \aap, 598, L5

\bibitem[{{Astropy Collaboration} {et~al.}(2018){Astropy Collaboration},
  {Price-Whelan}, {Sip{\H{o}}cz}, {G{\"u}nther}, {Lim}, {Crawford}, {Conseil},
  {Shupe}, {Craig}, {Dencheva}, {Ginsburg}, {Vand erPlas}, {Bradley},
  {P{\'e}rez-Su{\'a}rez}, {de Val-Borro}, {Aldcroft}, {Cruz}, {Robitaille},
  {Tollerud}, {Ardelean}, {Babej}, {Bach}, {Bachetti}, {Bakanov}, {Bamford},
  {Barentsen}, {Barmby}, {Baumbach}, {Berry}, {Biscani}, {Boquien}, {Bostroem},
  {Bouma}, {Brammer}, {Bray}, {Breytenbach}, {Buddelmeijer}, {Burke},
  {Calderone}, {Cano Rodr{\'\i}guez}, {Cara}, {Cardoso}, {Cheedella}, {Copin},
  {Corrales}, {Crichton}, {D'Avella}, {Deil}, {Depagne}, {Dietrich}, {Donath},
  {Droettboom}, {Earl}, {Erben}, {Fabbro}, {Ferreira}, {Finethy}, {Fox},
  {Garrison}, {Gibbons}, {Goldstein}, {Gommers}, {Greco}, {Greenfield},
  {Groener}, {Grollier}, {Hagen}, {Hirst}, {Homeier}, {Horton}, {Hosseinzadeh},
  {Hu}, {Hunkeler}, {Ivezi{\'c}}, {Jain}, {Jenness}, {Kanarek}, {Kendrew},
  {Kern}, {Kerzendorf}, {Khvalko}, {King}, {Kirkby}, {Kulkarni}, {Kumar},
  {Lee}, {Lenz}, {Littlefair}, {Ma}, {Macleod}, {Mastropietro}, {McCully},
  {Montagnac}, {Morris}, {Mueller}, {Mumford}, {Muna}, {Murphy}, {Nelson},
  {Nguyen}, {Ninan}, {N{\"o}the}, {Ogaz}, {Oh}, {Parejko}, {Parley}, {Pascual},
  {Patil}, {Patil}, {Plunkett}, {Prochaska}, {Rastogi}, {Reddy Janga},
  {Sabater}, {Sakurikar}, {Seifert}, {Sherbert}, {Sherwood-Taylor}, {Shih},
  {Sick}, {Silbiger}, {Singanamalla}, {Singer}, {Sladen}, {Sooley},
  {Sornarajah}, {Streicher}, {Teuben}, {Thomas}, {Tremblay}, {Turner},
  {Terr{\'o}n}, {van Kerkwijk}, {de la Vega}, {Watkins}, {Weaver}, {Whitmore},
  {Woillez}, {Zabalza}, \& {Astropy Contributors}}]{astropy:2018}
{Astropy Collaboration}, {Price-Whelan}, A.~M., {Sip{\H{o}}cz}, B.~M., {et~al.}
  2018, \aj, 156, 123

\bibitem[{{Astropy Collaboration} {et~al.}(2013){Astropy Collaboration},
  {Robitaille}, {Tollerud}, {Greenfield}, {Droettboom}, {Bray}, {Aldcroft},
  {Davis}, {Ginsburg}, {Price-Whelan}, {Kerzendorf}, {Conley}, {Crighton},
  {Barbary}, {Muna}, {Ferguson}, {Grollier}, {Parikh}, {Nair}, {Unther},
  {Deil}, {Woillez}, {Conseil}, {Kramer}, {Turner}, {Singer}, {Fox}, {Weaver},
  {Zabalza}, {Edwards}, {Azalee Bostroem}, {Burke}, {Casey}, {Crawford},
  {Dencheva}, {Ely}, {Jenness}, {Labrie}, {Lim}, {Pierfederici}, {Pontzen},
  {Ptak}, {Refsdal}, {Servillat}, \& {Streicher}}]{astropy:2013}
{Astropy Collaboration}, {Robitaille}, T.~P., {Tollerud}, E.~J., {et~al.} 2013,
  \aap, 558, A33

\bibitem[{{Barclay} {et~al.}(2018){Barclay}, {Pepper}, \&
  {Quintana}}]{barclay18}
{Barclay}, T., {Pepper}, J., \& {Quintana}, E.~V. 2018, \apjs, 239, 2

\bibitem[{{Bardalez Gagliuffi} {et~al.}(2019){Bardalez Gagliuffi}, {Burgasser},
  {Schmidt}, {Theissen}, {Gagn{\'e}}, {Gillon}, {Sahlmann}, {Faherty},
  {Gelino}, {Cruz}, {Skrzypek}, \& {Looper}}]{Gagliuffi2019}
{Bardalez Gagliuffi}, D.~C., {Burgasser}, A.~J., {Schmidt}, S.~J., {et~al.}
  2019, The Astrophysical Journal, 883, 205

\bibitem[{{Batalha} {et~al.}(2017){Batalha}, {Mandell}, {Pontoppidan},
  {Stevenson}, {Lewis}, {Kalirai}, {Earl}, {Greene}, {Albert}, \&
  {Nielsen}}]{Batalha2017}
{Batalha}, N.~E., {Mandell}, A., {Pontoppidan}, K., {et~al.} 2017, Publications
  of the Astronomical Society of the Pacific, 129, 064501

\bibitem[{{Benneke} {et~al.}(2019){Benneke}, {Wong}, {Piaulet}, {Knutson},
  {Lothringer}, {Morley}, {Crossfield}, {Gao}, {Greene}, {Dressing},
  {Dragomir}, {Howard}, {McCullough}, {Kempton}, {Fortney}, \&
  {Fraine}}]{benneke19}
{Benneke}, B., {Wong}, I., {Piaulet}, C., {et~al.} 2019, \apjl, 887, L14

\bibitem[{{Berta-Thompson} {et~al.}(2015){Berta-Thompson}, {Irwin},
  {Charbonneau}, {Newton}, {Dittmann}, {Astudillo-Defru}, {Bonfils}, {Gillon},
  {Jehin}, {Stark}, {Stalder}, {Bouchy}, {Delfosse}, {Forveille}, {Lovis},
  {Mayor}, {Neves}, {Pepe}, {Santos}, {Udry}, \& {W{\"u}nsche}}]{BertaT15}
{Berta-Thompson}, Z.~K., {Irwin}, J., {Charbonneau}, D., {et~al.} 2015, \nat,
  527, 204

\bibitem[{{Burdanov} {et~al.}(2018){Burdanov}, {Delrez}, {Gillon}, \&
  {Jehin}}]{Burdanov2018}
{Burdanov}, A., {Delrez}, L., {Gillon}, M., \& {Jehin}, E. 2018, {SPECULOOS
  Exoplanet Search and Its Prototype on TRAPPIST} (Springer International
  Publishing AG), 130

\bibitem[{{Burgasser} {et~al.}(2015){Burgasser}, {Gillon}, {Melis}, {Bowler},
  {Michelsen}, {Bardalez Gagliuffi}, {Gelino}, {Jehin}, {Delrez}, {Manfroid},
  \& {Blake}}]{Burgasser2015}
{Burgasser}, A.~J., {Gillon}, M., {Melis}, C., {et~al.} 2015, The Astronomical
  Journal, 149, 104

\bibitem[{{Charbonneau} {et~al.}(2009){Charbonneau}, {Berta}, {Irwin}, {Burke},
  {Nutzman}, {Buchhave}, {Lovis}, {Bonfils}, {Latham}, {Udry}, {Murray-Clay},
  {Holman}, {Falco}, {Winn}, {Queloz}, {Pepe}, {Mayor}, {Delfosse}, \&
  {Forveille}}]{cha09}
{Charbonneau}, D., {Berta}, Z.~K., {Irwin}, J., {et~al.} 2009, \nat, 462, 891

\bibitem[{{Coleman} {et~al.}(2019){Coleman}, {Leleu}, {Alibert}, \&
  {Benz}}]{coleman19}
{Coleman}, G.~A.~L., {Leleu}, A., {Alibert}, Y., \& {Benz}, W. 2019, \aap, 631,
  A7

\bibitem[{{Crossfield} {et~al.}(2019){Crossfield}, {Waalkes}, {Newton},
  {Narita}, {Muirhead}, {Ment}, {Matthews}, {Kraus}, {Kostov}, {Kosiarek},
  {Kane}, {Isaacson}, {Halverson}, {Gonzales}, {Everett}, {Dragomir},
  {Collins}, {Chontos}, {Berardo}, {Winters}, {Winn}, {Scott}, {Rojas-Ayala},
  {Rizzuto}, {Petigura}, {Peterson}, {Mocnik}, {Mikal-Evans}, {Mehrle},
  {Matson}, {Kuzuhara}, {Irwin}, {Huber}, {Huang}, {Howell}, {Howard},
  {Hirano}, {Fulton}, {Dupuy}, {Dressing}, {Dalba}, {Charbonneau}, {Burt},
  {Berta-Thompson}, {Benneke}, {Watanabe}, {Twicken}, {Tamura}, {Schlieder},
  {Seager}, {Rose}, {Ricker}, {Quintana}, {L{\'e}pine}, {Latham}, {Kotani},
  {Jenkins}, {Hori}, {Colon}, \& {Caldwell}}]{crossfield19}
{Crossfield}, I. J.~M., {Waalkes}, W., {Newton}, E.~R., {et~al.} 2019, \apjl,
  883, L16

\bibitem[{{Cruz} \& {Reid}(2002)}]{cruz02}
{Cruz}, K.~L. \& {Reid}, I.~N. 2002, \aj, 123, 2828

\bibitem[{{Cruz} {et~al.}(2003){Cruz}, {Reid}, {Liebert}, {Kirkpatrick}, \&
  {Lowrance}}]{cruz03}
{Cruz}, K.~L., {Reid}, I.~N., {Liebert}, J., {Kirkpatrick}, J.~D., \&
  {Lowrance}, P.~J. 2003, \aj, 126, 2421

\bibitem[{{de Wit} \& {Seager}(2013)}]{dewit2013}
{de Wit}, J. \& {Seager}, S. 2013, Science, 342, 1473

\bibitem[{{Delrez} {et~al.}(2018){Delrez}, {Gillon}, {Queloz}, {Demory},
  {Almleaky}, {de Wit}, {Jehin}, {Triaud}, {Barkaoui}, {Burdanov}, {Burgasser},
  {Ducrot}, {McCormac}, {Murray}, {Silva Fernandes}, {Sohy}, {Thompson}, {Van
  Grootel}, {Alonso}, {Benkhaldoun}, \& {Rebolo}}]{Delrez2018b}
{Delrez}, L., {Gillon}, M., {Queloz}, D., {et~al.} 2018, in Society of
  Photo-Optical Instrumentation Engineers (SPIE) Conference Series, Vol. 10700,
  Proceedings of the SPIE, 107001I

\bibitem[{Delrez {et~al.}(2018)Delrez, Gillon, Triaud, Demory, de~Wit, Ingalls,
  Agol, Bolmont, Burdanov, Burgasser, Carey, Jehin, Leconte, Lederer, Queloz,
  Selsis, \& Grootel}]{Delrez2018a}
Delrez, L., Gillon, M., Triaud, A. H. M.~J., {et~al.} 2018, Monthly Notices of
  the Royal Astronomical Society, 475, 3577

\bibitem[{{Demory} {et~al.}(2016){Demory}, {Gillon}, {Madhusudhan}, \&
  {Queloz}}]{Demory2016}
{Demory}, B.-O., {Gillon}, M., {Madhusudhan}, N., \& {Queloz}, D. 2016, Monthly
  Notices of the Royal Astronomical Society, 455, 2018

\bibitem[{Demory(2020)}]{demory2020}
Demory, B.-O. e.~a. 2020, Astronomy {\&} Astrophysics, in press

\bibitem[{{Dieterich} {et~al.}(2014){Dieterich}, {Henry}, {Jao}, {Winters},
  {Hosey}, {Riedel}, \& {Subasavage}}]{Dieterich2014}
{Dieterich}, S.~B., {Henry}, T.~J., {Jao}, W.-C., {et~al.} 2014, The
  Astronomical Journal, 147, 94

\bibitem[{{Dittmann} {et~al.}(2017){Dittmann}, {Irwin}, {Charbonneau},
  {Bonfils}, {Astudillo-Defru}, {Haywood}, {Berta-Thompson}, {Newton},
  {Rodriguez}, {Winters}, {Tan}, {Almenara}, {Bouchy}, {Delfosse}, {Forveille},
  {Lovis}, {Murgas}, {Pepe}, {Santos}, {Udry}, {W{\"u}nsche}, {Esquerdo},
  {Latham}, \& {Dressing}}]{dittmann17}
{Dittmann}, J.~A., {Irwin}, J.~M., {Charbonneau}, D., {et~al.} 2017, \nat, 544,
  333

\bibitem[{{Filippazzo} {et~al.}(2015){Filippazzo}, {Rice}, {Faherty}, {Cruz},
  {Van Gordon}, \& {Looper}}]{F2015}
{Filippazzo}, J.~C., {Rice}, E.~L., {Faherty}, J., {et~al.} 2015, The
  Astrophysical Journal, 810, 158

\bibitem[{{Gaia Collaboration} {et~al.}(2018){Gaia Collaboration}, {Brown},
  {Vallenari}, {Prusti}, {de Bruijne}, {Babusiaux}, {Bailer-Jones}, {Biermann},
  {Evans}, {Eyer}, {Jansen}, {Jordi}, {Klioner}, {Lammers}, {Lindegren},
  {Luri}, {Mignard}, {Panem}, {Pourbaix}, {Randich}, {Sartoretti}, {Siddiqui},
  {Soubiran}, {van Leeuwen}, {Walton}, {Arenou}, {Bastian}, {Cropper},
  {Drimmel}, {Katz}, {Lattanzi}, {Bakker}, {Cacciari}, {Casta{\~n}eda},
  {Chaoul}, {Cheek}, {De Angeli}, {Fabricius}, {Guerra}, {Holl}, {Masana},
  {Messineo}, {Mowlavi}, {Nienartowicz}, {Panuzzo}, {Portell}, {Riello},
  {Seabroke}, {Tanga}, {Th{\'e}venin}, {Gracia-Abril}, {Comoretto},
  {Garcia-Reinaldos}, {Teyssier}, {Altmann}, {Andrae}, {Audard},
  {Bellas-Velidis}, {Benson}, {Berthier}, {Blomme}, {Burgess}, {Busso},
  {Carry}, {Cellino}, {Clementini}, {Clotet}, {Creevey}, {Davidson}, {De
  Ridder}, {Delchambre}, {Dell'Oro}, {Ducourant},
  {Fern{\'a}ndez-Hern{\'a}ndez}, {Fouesneau}, {Fr{\'e}mat}, {Galluccio},
  {Garc{\'\i}a-Torres}, {Gonz{\'a}lez-N{\'u}{\~n}ez}, {Gonz{\'a}lez-Vidal},
  {Gosset}, {Guy}, {Halbwachs}, {Hambly}, {Harrison}, {Hern{\'a}ndez},
  {Hestroffer}, {Hodgkin}, {Hutton}, {Jasniewicz}, {Jean-Antoine-Piccolo},
  {Jordan}, {Korn}, {Krone-Martins}, {Lanzafame}, {Lebzelter}, {L{\"o}ffler},
  {Manteiga}, {Marrese}, {Mart{\'\i}n-Fleitas}, {Moitinho}, {Mora}, {Muinonen},
  {Osinde}, {Pancino}, {Pauwels}, {Petit}, {Recio-Blanco}, {Richards},
  {Rimoldini}, {Robin}, {Sarro}, {Siopis}, {Smith}, {Sozzetti}, {S{\"u}veges},
  {Torra}, {van Reeven}, {Abbas}, {Abreu Aramburu}, {Accart}, {Aerts},
  {Altavilla}, {{\'A}lvarez}, {Alvarez}, {Alves}, {Anderson}, {Andrei},
  {Anglada Varela}, {Antiche}, {Antoja}, {Arcay}, {Astraatmadja}, {Bach},
  {Baker}, {Balaguer-N{\'u}{\~n}ez}, {Balm}, {Barache}, {Barata}, {Barbato},
  {Barblan}, {Barklem}, {Barrado}, {Barros}, {Barstow}, {Bartholom{\'e}
  Mu{\~n}oz}, {Bassilana}, {Becciani}, {Bellazzini}, {Berihuete}, {Bertone},
  {Bianchi}, {Bienaym{\'e}}, {Blanco-Cuaresma}, {Boch}, {Boeche}, {Bombrun},
  {Borrachero}, {Bossini}, {Bouquillon}, {Bourda}, {Bragaglia}, {Bramante},
  {Breddels}, {Bressan}, {Brouillet}, {Br{\"u}semeister}, {Brugaletta},
  {Bucciarelli}, {Burlacu}, {Busonero}, {Butkevich}, {Buzzi}, {Caffau},
  {Cancelliere}, {Cannizzaro}, {Cantat-Gaudin}, {Carballo}, {Carlucci},
  {Carrasco}, {Casamiquela}, {Castellani}, {Castro-Ginard}, {Charlot},
  {Chemin}, {Chiavassa}, {Cocozza}, {Costigan}, {Cowell}, {Crifo}, {Crosta},
  {Crowley}, {Cuypers}, {Dafonte}, {Damerdji}, {Dapergolas}, {David}, {David},
  {de Laverny}, {De Luise}, {De March}, {de Martino}, {de Souza}, {de Torres},
  {Debosscher}, {del Pozo}, {Delbo}, {Delgado}, {Delgado}, {Di Matteo},
  {Diakite}, {Diener}, {Distefano}, {Dolding}, {Drazinos}, {Dur{\'a}n},
  {Edvardsson}, {Enke}, {Eriksson}, {Esquej}, {Eynard Bontemps}, {Fabre},
  {Fabrizio}, {Faigler}, {Falc{\~a}o}, {Farr{\`a}s Casas}, {Federici},
  {Fedorets}, {Fernique}, {Figueras}, {Filippi}, {Findeisen}, {Fonti},
  {Fraile}, {Fraser}, {Fr{\'e}zouls}, {Gai}, {Galleti}, {Garabato},
  {Garc{\'\i}a-Sedano}, {Garofalo}, {Garralda}, {Gavel}, {Gavras}, {Gerssen},
  {Geyer}, {Giacobbe}, {Gilmore}, {Girona}, {Giuffrida}, {Glass}, {Gomes},
  {Granvik}, {Gueguen}, {Guerrier}, {Guiraud}, {Guti{\'e}rrez-S{\'a}nchez},
  {Haigron}, {Hatzidimitriou}, {Hauser}, {Haywood}, {Heiter}, {Helmi}, {Heu},
  {Hilger}, {Hobbs}, {Hofmann}, {Holland}, {Huckle}, {Hypki}, {Icardi},
  {Jan{\ss}en}, {Jevardat de Fombelle}, {Jonker}, {Juh{\'a}sz}, {Julbe},
  {Karampelas}, {Kewley}, {Klar}, {Kochoska}, {Kohley}, {Kolenberg},
  {Kontizas}, {Kontizas}, {Koposov}, {Kordopatis}, {Kostrzewa-Rutkowska},
  {Koubsky}, {Lambert}, {Lanza}, {Lasne}, {Lavigne}, {Le Fustec}, {Le
  Poncin-Lafitte}, {Lebreton}, {Leccia}, {Leclerc}, {Lecoeur-Taibi},
  {Lenhardt}, {Leroux}, {Liao}, {Licata}, {Lindstr{\o}m}, {Lister}, {Livanou},
  {Lobel}, {L{\'o}pez}, {Managau}, {Mann}, {Mantelet}, {Marchal}, {Marchant},
  {Marconi}, {Marinoni}, {Marschalk{\'o}}, {Marshall}, {Martino}, {Marton},
  {Mary}, {Massari}, {Matijevi{\v{c}}}, {Mazeh}, {McMillan}, {Messina},
  {Michalik}, {Millar}, {Molina}, {Molinaro}, {Moln{\'a}r}, {Montegriffo},
  {Mor}, {Morbidelli}, {Morel}, {Morris}, {Mulone}, {Muraveva}, {Musella},
  {Nelemans}, {Nicastro}, {Noval}, {O'Mullane}, {Ord{\'e}novic},
  {Ord{\'o}{\~n}ez-Blanco}, {Osborne}, {Pagani}, {Pagano}, {Pailler},
  {Palacin}, {Palaversa}, {Panahi}, {Pawlak}, {Piersimoni}, {Pineau}, {Plachy},
  {Plum}, {Poggio}, {Poujoulet}, {Pr{\v{s}}a}, {Pulone}, {Racero}, {Ragaini},
  {Rambaux}, {Ramos-Lerate}, {Regibo}, {Reyl{\'e}}, {Riclet}, {Ripepi}, {Riva},
  {Rivard}, {Rixon}, {Roegiers}, {Roelens}, {Romero-G{\'o}mez}, {Rowell},
  {Royer}, {Ruiz-Dern}, {Sadowski}, {Sagrist{\`a} Sell{\'e}s}, {Sahlmann},
  {Salgado}, {Salguero}, {Sanna}, {Santana-Ros}, {Sarasso}, {Savietto},
  {Schultheis}, {Sciacca}, {Segol}, {Segovia}, {S{\'e}gransan}, {Shih},
  {Siltala}, {Silva}, {Smart}, {Smith}, {Solano}, {Solitro}, {Sordo}, {Soria
  Nieto}, {Souchay}, {Spagna}, {Spoto}, {Stampa}, {Steele},
  {Steidelm{\"u}ller}, {Stephenson}, {Stoev}, {Suess}, {Surdej}, {Szabados},
  {Szegedi-Elek}, {Tapiador}, {Taris}, {Tauran}, {Taylor}, {Teixeira},
  {Terrett}, {Teyssand ier}, {Thuillot}, {Titarenko}, {Torra Clotet}, {Turon},
  {Ulla}, {Utrilla}, {Uzzi}, {Vaillant}, {Valentini}, {Valette}, {van Elteren},
  {Van Hemelryck}, {van Leeuwen}, {Vaschetto}, {Vecchiato}, {Veljanoski},
  {Viala}, {Vicente}, {Vogt}, {von Essen}, {Voss}, {Votruba}, {Voutsinas},
  {Walmsley}, {Weiler}, {Wertz}, {Wevers}, {Wyrzykowski}, {Yoldas},
  {{\v{Z}}erjal}, {Ziaeepour}, {Zorec}, {Zschocke}, {Zucker}, {Zurbach}, \&
  {Zwitter}}]{Gaia2018}
{Gaia Collaboration}, {Brown}, A.~G.~A., {Vallenari}, A., {et~al.} 2018,
  Astronomy and Astrophysics, 616, A1

\bibitem[{{Gaia Collaboration} {et~al.}(2016){Gaia Collaboration}, {Prusti},
  {de Bruijne}, {Brown}, {Vallenari}, {Babusiaux}, {Bailer-Jones}, {Bastian},
  {Biermann}, {Evans}, {Eyer}, {Jansen}, {Jordi}, {Klioner}, {Lammers},
  {Lindegren}, {Luri}, {Mignard}, {Milligan}, {Panem}, {Poinsignon},
  {Pourbaix}, {Randich}, {Sarri}, {Sartoretti}, {Siddiqui}, {Soubiran},
  {Valette}, {van Leeuwen}, {Walton}, {Aerts}, {Arenou}, {Cropper}, {Drimmel},
  {H{\o}g}, {Katz}, {Lattanzi}, {O'Mullane}, {Grebel}, {Holland}, {Huc},
  {Passot}, {Bramante}, {Cacciari}, {Casta{\~n}eda}, {Chaoul}, {Cheek}, {De
  Angeli}, {Fabricius}, {Guerra}, {Hern{\'a}ndez}, {Jean-Antoine-Piccolo},
  {Masana}, {Messineo}, {Mowlavi}, {Nienartowicz}, {Ord{\'o}{\~n}ez-Blanco},
  {Panuzzo}, {Portell}, {Richards}, {Riello}, {Seabroke}, {Tanga},
  {Th{\'e}venin}, {Torra}, {Els}, {Gracia-Abril}, {Comoretto},
  {Garcia-Reinaldos}, {Lock}, {Mercier}, {Altmann}, {Andrae}, {Astraatmadja},
  {Bellas-Velidis}, {Benson}, {Berthier}, {Blomme}, {Busso}, {Carry},
  {Cellino}, {Clementini}, {Cowell}, {Creevey}, {Cuypers}, {Davidson}, {De
  Ridder}, {de Torres}, {Delchambre}, {Dell'Oro}, {Ducourant}, {Fr{\'e}mat},
  {Garc{\'\i}a-Torres}, {Gosset}, {Halbwachs}, {Hambly}, {Harrison}, {Hauser},
  {Hestroffer}, {Hodgkin}, {Huckle}, {Hutton}, {Jasniewicz}, {Jordan},
  {Kontizas}, {Korn}, {Lanzafame}, {Manteiga}, {Moitinho}, {Muinonen},
  {Osinde}, {Pancino}, {Pauwels}, {Petit}, {Recio-Blanco}, {Robin}, {Sarro},
  {Siopis}, {Smith}, {Smith}, {Sozzetti}, {Thuillot}, {van Reeven}, {Viala},
  {Abbas}, {Abreu Aramburu}, {Accart}, {Aguado}, {Allan}, {Allasia},
  {Altavilla}, {{\'A}lvarez}, {Alves}, {Anderson}, {Andrei}, {Anglada Varela},
  {Antiche}, {Antoja}, {Ant{\'o}n}, {Arcay}, {Atzei}, {Ayache}, {Bach},
  {Baker}, {Balaguer-N{\'u}{\~n}ez}, {Barache}, {Barata}, {Barbier}, {Barblan},
  {Baroni}, {Barrado y Navascu{\'e}s}, {Barros}, {Barstow}, {Becciani},
  {Bellazzini}, {Bellei}, {Bello Garc{\'\i}a}, {Belokurov}, {Bendjoya},
  {Berihuete}, {Bianchi}, {Bienaym{\'e}}, {Billebaud}, {Blagorodnova},
  {Blanco-Cuaresma}, {Boch}, {Bombrun}, {Borrachero}, {Bouquillon}, {Bourda},
  {Bouy}, {Bragaglia}, {Breddels}, {Brouillet}, {Br{\"u}semeister},
  {Bucciarelli}, {Budnik}, {Burgess}, {Burgon}, {Burlacu}, {Busonero}, {Buzzi},
  {Caffau}, {Cambras}, {Campbell}, {Cancelliere}, {Cantat-Gaudin}, {Carlucci},
  {Carrasco}, {Castellani}, {Charlot}, {Charnas}, {Charvet}, {Chassat},
  {Chiavassa}, {Clotet}, {Cocozza}, {Collins}, {Collins}, {Costigan}, {Crifo},
  {Cross}, {Crosta}, {Crowley}, {Dafonte}, {Damerdji}, {Dapergolas}, {David},
  {David}, {De Cat}, {de Felice}, {de Laverny}, {De Luise}, {De March}, {de
  Martino}, {de Souza}, {Debosscher}, {del Pozo}, {Delbo}, {Delgado},
  {Delgado}, {di Marco}, {Di Matteo}, {Diakite}, {Distefano}, {Dolding}, {Dos
  Anjos}, {Drazinos}, {Dur{\'a}n}, {Dzigan}, {Ecale}, {Edvardsson}, {Enke},
  {Erdmann}, {Escolar}, {Espina}, {Evans}, {Eynard Bontemps}, {Fabre},
  {Fabrizio}, {Faigler}, {Falc{\~a}o}, {Farr{\`a}s Casas}, {Faye}, {Federici},
  {Fedorets}, {Fern{\'a}ndez-Hern{\'a}ndez}, {Fernique}, {Fienga}, {Figueras},
  {Filippi}, {Findeisen}, {Fonti}, {Fouesneau}, {Fraile}, {Fraser}, {Fuchs},
  {Furnell}, {Gai}, {Galleti}, {Galluccio}, {Garabato}, {Garc{\'\i}a-Sedano},
  {Gar{\'e}}, {Garofalo}, {Garralda}, {Gavras}, {Gerssen}, {Geyer}, {Gilmore},
  {Girona}, {Giuffrida}, {Gomes}, {Gonz{\'a}lez-Marcos},
  {Gonz{\'a}lez-N{\'u}{\~n}ez}, {Gonz{\'a}lez-Vidal}, {Granvik}, {Guerrier},
  {Guillout}, {Guiraud}, {G{\'u}rpide}, {Guti{\'e}rrez-S{\'a}nchez}, {Guy},
  {Haigron}, {Hatzidimitriou}, {Haywood}, {Heiter}, {Helmi}, {Hobbs},
  {Hofmann}, {Holl}, {Holland }, {Hunt}, {Hypki}, {Icardi}, {Irwin}, {Jevardat
  de Fombelle}, {Jofr{\'e}}, {Jonker}, {Jorissen}, {Julbe}, {Karampelas},
  {Kochoska}, {Kohley}, {Kolenberg}, {Kontizas}, {Koposov}, {Kordopatis},
  {Koubsky}, {Kowalczyk}, {Krone-Martins}, {Kudryashova}, {Kull}, {Bachchan},
  {Lacoste-Seris}, {Lanza}, {Lavigne}, {Le Poncin-Lafitte}, {Lebreton},
  {Lebzelter}, {Leccia}, {Leclerc}, {Lecoeur-Taibi}, {Lemaitre}, {Lenhardt},
  {Leroux}, {Liao}, {Licata}, {Lindstr{\o}m}, {Lister}, {Livanou}, {Lobel},
  {L{\"o}ffler}, {L{\'o}pez}, {Lopez-Lozano}, {Lorenz}, {Loureiro},
  {MacDonald}, {Magalh{\~a}es Fernandes}, {Managau}, {Mann}, {Mantelet},
  {Marchal}, {Marchant}, {Marconi}, {Marie}, {Marinoni}, {Marrese},
  {Marschalk{\'o}}, {Marshall}, {Mart{\'\i}n-Fleitas}, {Martino}, {Mary},
  {Matijevi{\v{c}}}, {Mazeh}, {McMillan}, {Messina}, {Mestre}, {Michalik},
  {Millar}, {Miranda}, {Molina}, {Molinaro}, {Molinaro}, {Moln{\'a}r},
  {Moniez}, {Montegriffo}, {Monteiro}, {Mor}, {Mora}, {Morbidelli}, {Morel},
  {Morgenthaler}, {Morley}, {Morris}, {Mulone}, {Muraveva}, {Musella},
  {Narbonne}, {Nelemans}, {Nicastro}, {Noval}, {Ord{\'e}novic},
  {Ordieres-Mer{\'e}}, {Osborne}, {Pagani}, {Pagano}, {Pailler}, {Palacin},
  {Palaversa}, {Parsons}, {Paulsen}, {Pecoraro}, {Pedrosa}, {Pentik{\"a}inen},
  {Pereira}, {Pichon}, {Piersimoni}, {Pineau}, {Plachy}, {Plum}, {Poujoulet},
  {Pr{\v{s}}a}, {Pulone}, {Ragaini}, {Rago}, {Rambaux}, {Ramos-Lerate},
  {Ranalli}, {Rauw}, {Read}, {Regibo}, {Renk}, {Reyl{\'e}}, {Ribeiro},
  {Rimoldini}, {Ripepi}, {Riva}, {Rixon}, {Roelens}, {Romero-G{\'o}mez},
  {Rowell}, {Royer}, {Rudolph}, {Ruiz-Dern}, {Sadowski}, {Sagrist{\`a}
  Sell{\'e}s}, {Sahlmann}, {Salgado}, {Salguero}, {Sarasso}, {Savietto},
  {Schnorhk}, {Schultheis}, {Sciacca}, {Segol}, {Segovia}, {Segransan},
  {Serpell}, {Shih}, {Smareglia}, {Smart}, {Smith}, {Solano}, {Solitro},
  {Sordo}, {Soria Nieto}, {Souchay}, {Spagna}, {Spoto}, {Stampa}, {Steele},
  {Steidelm{\"u}ller}, {Stephenson}, {Stoev}, {Suess}, {S{\"u}veges}, {Surdej},
  {Szabados}, {Szegedi-Elek}, {Tapiador}, {Taris}, {Tauran}, {Taylor},
  {Teixeira}, {Terrett}, {Tingley}, {Trager}, {Turon}, {Ulla}, {Utrilla},
  {Valentini}, {van Elteren}, {Van Hemelryck}, {van Leeuwen}, {Varadi},
  {Vecchiato}, {Veljanoski}, {Via}, {Vicente}, {Vogt}, {Voss}, {Votruba},
  {Voutsinas}, {Walmsley}, {Weiler}, {Weingrill}, {Werner}, {Wevers},
  {Whitehead}, {Wyrzykowski}, {Yoldas}, {{\v{Z}}erjal}, {Zucker}, {Zurbach},
  {Zwitter}, {Alecu}, {Allen}, {Allende Prieto}, {Amorim},
  {Anglada-Escud{\'e}}, {Arsenijevic}, {Azaz}, {Balm}, {Beck}, {Bernstein},
  {Bigot}, {Bijaoui}, {Blasco}, {Bonfigli}, {Bono}, {Boudreault}, {Bressan},
  {Brown}, {Brunet}, {Bunclark}, {Buonanno}, {Butkevich}, {Carret}, {Carrion},
  {Chemin}, {Ch{\'e}reau}, {Corcione}, {Darmigny}, {de Boer}, {de Teodoro}, {de
  Zeeuw}, {Delle Luche}, {Domingues}, {Dubath}, {Fodor}, {Fr{\'e}zouls},
  {Fries}, {Fustes}, {Fyfe}, {Gallardo}, {Gallegos}, {Gardiol}, {Gebran},
  {Gomboc}, {G{\'o}mez}, {Grux}, {Gueguen}, {Heyrovsky}, {Hoar}, {Iannicola},
  {Isasi Parache}, {Janotto}, {Joliet}, {Jonckheere}, {Keil}, {Kim},
  {Klagyivik}, {Klar}, {Knude}, {Kochukhov}, {Kolka}, {Kos}, {Kutka}, {Lainey},
  {LeBouquin}, {Liu}, {Loreggia}, {Makarov}, {Marseille}, {Martayan},
  {Martinez-Rubi}, {Massart}, {Meynadier}, {Mignot}, {Munari}, {Nguyen},
  {Nordlander}, {Ocvirk}, {O'Flaherty}, {Olias Sanz}, {Ortiz}, {Osorio},
  {Oszkiewicz}, {Ouzounis}, {Palmer}, {Park}, {Pasquato}, {Peltzer}, {Peralta},
  {P{\'e}turaud}, {Pieniluoma}, {Pigozzi}, {Poels}, {Prat}, {Prod'homme},
  {Raison}, {Rebordao}, {Risquez}, {Rocca-Volmerange}, {Rosen}, {Ruiz-Fuertes},
  {Russo}, {Sembay}, {Serraller Vizcaino}, {Short}, {Siebert}, {Silva},
  {Sinachopoulos}, {Slezak}, {Soffel}, {Sosnowska}, {Strai{\v{z}}ys}, {ter
  Linden}, {Terrell}, {Theil}, {Tiede}, {Troisi}, {Tsalmantza}, {Tur},
  {Vaccari}, {Vachier}, {Valles}, {Van Hamme}, {Veltz}, {Virtanen}, {Wallut},
  {Wichmann}, {Wilkinson}, {Ziaeepour}, \& {Zschocke}}]{Gaia2016}
{Gaia Collaboration}, {Prusti}, T., {de Bruijne}, J.~H.~J., {et~al.} 2016,
  Astronomy and Astrophysics, 595, A1

\bibitem[{{Gibbs} {et~al.}(2020){Gibbs}, {Bixel}, {Rackham}, {Apai},
  {Schlecker}, {Espinoza}, {Mancini}, {Chen}, {Henning}, {Gabor}, {Boyle},
  {Perez Chavez}, {Mousseau}, {Dietrich}, {Jay Socia}, {Ip}, {Ngeow}, {Tsai},
  {Bhand are}, {Marian}, {Baehr}, {Brown}, {H{\"a}berle}, {Keppler},
  {Molaverdikhani}, \& {Sarkis}}]{gibbs20}
{Gibbs}, A., {Bixel}, A., {Rackham}, B.~V., {et~al.} 2020, \aj, 159, 169

\bibitem[{{Gillon}(2018)}]{Gillon2018}
{Gillon}, M. 2018, Nature Astronomy, 2, 344

\bibitem[{{Gillon} {et~al.}(2013){Gillon}, {Jehin}, {Fumel}, {Magain}, \&
  {Queloz}}]{Gillon2013}
{Gillon}, M., {Jehin}, E., {Fumel}, A., {Magain}, P., \& {Queloz}, D. 2013, in
  European Physical Journal Web of Conferences, Vol.~47, European Physical
  Journal Web of Conferences, 03001

\bibitem[{{Gillon} {et~al.}(2016){Gillon}, {Jehin}, {Lederer}, {Delrez}, {de
  Wit}, {Burdanov}, {Van Grootel}, {Burgasser}, {Triaud}, {Opitom}, {Demory},
  {Sahu}, {Bardalez Gagliuffi}, {Magain}, \& {Queloz}}]{Gillon2016}
{Gillon}, M., {Jehin}, E., {Lederer}, S.~M., {et~al.} 2016, Nature, 533, 221

\bibitem[{{Gillon} {et~al.}(2011){Gillon}, {Jehin}, {Magain}, {Chantry},
  {Hutsem{\'e}kers}, {Manfroid}, {Queloz}, \& {Udry}}]{Gillon11}
{Gillon}, M., {Jehin}, E., {Magain}, P., {et~al.} 2011, in European Physical
  Journal Web of Conferences, Vol.~11, European Physical Journal Web of
  Conferences, 06002

\bibitem[{{Gillon} {et~al.}(2020){Gillon}, {Meadows}, {Agol}, {Burgasser},
  {Deming}, {Doyon}, {Fortney}, {Kreidberg}, {Owen}, {Selsis}, {de Wit},
  {Lustig-Yaeger}, \& {Rackham}}]{gillon20}
{Gillon}, M., {Meadows}, V., {Agol}, E., {et~al.} 2020, arXiv e-prints,
  arXiv:2002.04798

\bibitem[{Gillon {et~al.}(2017)Gillon, Triaud, Demory, Jehin, Agol, Deck,
  Lederer, de~Wit, Burdanov, Ingalls, Bolmont, Leconte, Raymond, Selsis,
  Turbet, Barkaoui, Burgasser, Burleigh, Carey, Chaushev, Copperwheat, Delrez,
  Fernandes, Holdsworth, Kotze, Grootel, Almleaky, Benkhaldoun, Magain, \&
  Queloz}]{Gillon2017}
Gillon, M., Triaud, A. H. M.~J., Demory, B.-O., {et~al.} 2017, Nature, 542, 456

\bibitem[{{Gizis}(1997)}]{gizis97}
{Gizis}, J.~E. 1997, \aj, 113, 806

\bibitem[{{G{\"u}nther} {et~al.}(2019){G{\"u}nther}, {Pozuelos}, {Dittmann},
  {Dragomir}, {Kane}, {Daylan}, {Feinstein}, {Huang}, {Morton}, {Bonfanti},
  {Bouma}, {Burt}, {Collins}, {Lissauer}, {Matthews}, {Montet}, {Vand erburg},
  {Wang}, {Winters}, {Ricker}, {Vanderspek}, {Latham}, {Seager}, {Winn},
  {Jenkins}, {Armstrong}, {Barkaoui}, {Batalha}, {Bean}, {Caldwell}, {Ciardi},
  {Collins}, {Crossfield}, {Fausnaugh}, {Furesz}, {Gan}, {Gillon}, {Guerrero},
  {Horne}, {Howell}, {Ireland }, {Isopi}, {Jehin}, {Kielkopf}, {Lepine},
  {Mallia}, {Matson}, {Myers}, {Palle}, {Quinn}, {Relles}, {Rojas-Ayala},
  {Schlieder}, {Sefako}, {Shporer}, {Su{\'a}rez}, {Tan}, {Ting}, {Twicken}, \&
  {Waite}}]{gunther2019}
{G{\"u}nther}, M.~N., {Pozuelos}, F.~J., {Dittmann}, J.~A., {et~al.} 2019,
  Nature Astronomy, 3, 1099

\bibitem[{{Hardegree-Ullman} {et~al.}(2019){Hardegree-Ullman}, {Cushing},
  {Muirhead}, \& {Christiansen}}]{Ullman19}
{Hardegree-Ullman}, K.~K., {Cushing}, M.~C., {Muirhead}, P.~S., \&
  {Christiansen}, J.~L. 2019, \aj, 158, 75

\bibitem[{{Hawley} {et~al.}(2002){Hawley}, {Covey}, {Knapp}, {Golimowski},
  {Fan}, {Anderson}, {Gunn}, {Harris}, {Ivezi{\'c}}, {Long}, {Lupton},
  {McGehee}, {Narayanan}, {Peng}, {Schlegel}, {Schneider}, {Spahn}, {Strauss},
  {Szkody}, {Tsvetanov}, {Walkowicz}, {Brinkmann}, {Harvanek}, {Hennessy},
  {Kleinman}, {Krzesinski}, {Long}, {Neilsen}, {Newman}, {Nitta}, {Snedden}, \&
  {York}}]{hawley02}
{Hawley}, S.~L., {Covey}, K.~R., {Knapp}, G.~R., {et~al.} 2002, \aj, 123, 3409

\bibitem[{{He} {et~al.}(2017){He}, {Triaud}, \& {Gillon}}]{he17}
{He}, M.~Y., {Triaud}, A. H.~M.~J., \& {Gillon}, M. 2017, \mnras, 464, 2687

\bibitem[{Heller {et~al.}(2019)Heller, Hippke, \&
  Rodenbeck}]{Heller2019TransitK2}
Heller, R., Hippke, M., \& Rodenbeck, K. 2019, Astronomy {\&} Astrophysics,
  627, A66

\bibitem[{{Henry} {et~al.}(2018){Henry}, {Jao}, {Winters}, {Dieterich},
  {Finch}, {Ianna}, {Riedel}, {Silverstein}, {Subasavage}, \&
  {Vrijmoet}}]{Henry2018}
{Henry}, T.~J., {Jao}, W.-C., {Winters}, J.~G., {et~al.} 2018, The Astronomical
  Journal, 155, 265

\bibitem[{{Henry} {et~al.}(2004){Henry}, {Subasavage}, {Brown}, {Beaulieu},
  {Jao}, \& {Hambly}}]{Henry2004}
{Henry}, T.~J., {Subasavage}, J.~P., {Brown}, M.~A., {et~al.} 2004, The
  Astronomical Journal, 128, 2460

\bibitem[{Hippke {et~al.}(2019)Hippke, David, Mulders, \&
  Heller}]{Hippke2019Python}
Hippke, M., David, T.~J., Mulders, G.~D., \& Heller, R. 2019, The Astronomical
  Journal, 158, 143

\bibitem[{Hippke \& Heller(2019)}]{Hippke2019TransitPlanets}
Hippke, M. \& Heller, R. 2019, Astronomy and Astrophysics, 623, A39

\bibitem[{{Howard} {et~al.}(2012){Howard}, {Marcy}, {Bryson}, {Jenkins},
  {Rowe}, {Batalha}, {Borucki}, {Koch}, {Dunham}, {Gautier}, {Van Cleve},
  {Cochran}, {Latham}, {Lissauer}, {Torres}, {Brown}, {Gilliland}, {Buchhave},
  {Caldwell}, {Christensen-Dalsgaard}, {Ciardi}, {Fressin}, {Haas}, {Howell},
  {Kjeldsen}, {Seager}, {Rogers}, {Sasselov}, {Steffen}, {Basri},
  {Charbonneau}, {Christiansen}, {Clarke}, {Dupree}, {Fabrycky}, {Fischer},
  {Ford}, {Fortney}, {Tarter}, {Girouard}, {Holman}, {Johnson}, {Klaus},
  {Machalek}, {Moorhead}, {Morehead}, {Ragozzine}, {Tenenbaum}, {Twicken},
  {Quinn}, {Isaacson}, {Shporer}, {Lucas}, {Walkowicz}, {Welsh}, {Boss},
  {Devore}, {Gould}, {Smith}, {Morris}, {Prsa}, {Morton}, {Still}, {Thompson},
  {Mullally}, {Endl}, \& {MacQueen}}]{howard12}
{Howard}, A.~W., {Marcy}, G.~W., {Bryson}, S.~T., {et~al.} 2012, \apjs, 201, 15

\bibitem[{{Jehin} {et~al.}(2018){Jehin}, {Gillon}, {Queloz}, {Delrez},
  {Burdanov}, {Murray}, {Sohy}, {Ducrot}, {Sebastian}, {Thompson}, {McCormac},
  {Almleaky}, {Burgasser}, {Demory}, {de Wit}, {Barkaoui}, {Pozuelos},
  {Triaud}, \& {Grootel}}]{Jehin2018}
{Jehin}, E., {Gillon}, M., {Queloz}, D., {et~al.} 2018, The Messenger, 174, 2

\bibitem[{{Jehin} {et~al.}(2011){Jehin}, {Gillon}, {Queloz}, {Magain},
  {Manfroid}, {Chantry}, {Lendl}, {Hutsem{\'e}kers}, \& {Udry}}]{jehin11}
{Jehin}, E., {Gillon}, M., {Queloz}, D., {et~al.} 2011, The Messenger, 145, 2

\bibitem[{{Jenkins} {et~al.}(2016){Jenkins}, {Twicken}, {McCauliff},
  {Campbell}, {Sanderfer}, {Lung}, {Mansouri-Samani}, {Girouard}, {Tenenbaum},
  {Klaus}, {Smith}, {Caldwell}, {Chacon}, {Henze}, {Heiges}, {Latham},
  {Morgan}, {Swade}, {Rinehart}, \& {Vanderspek}}]{jenkins16}
{Jenkins}, J.~M., {Twicken}, J.~D., {McCauliff}, S., {et~al.} 2016, in Society
  of Photo-Optical Instrumentation Engineers (SPIE) Conference Series, Vol.
  9913, \procspie, 99133E

\bibitem[{{Jensen}(2013)}]{jensen2013}
{Jensen}, E. 2013, {Tapir: A web interface for transit/eclipse observability}

\bibitem[{{Kaltenegger} \& {Traub}(2009)}]{Lisa2009}
{Kaltenegger}, L. \& {Traub}, W.~A. 2009, The Astrophysical Journal, 698, 519

\bibitem[{{Kiman} {et~al.}(2019){Kiman}, {Schmidt}, {Angus}, {Cruz}, {Faherty},
  \& {Rice}}]{kiman19}
{Kiman}, R., {Schmidt}, S.~J., {Angus}, R., {et~al.} 2019, \aj, 157, 231

\bibitem[{{Kirkpatrick}(2005)}]{Kirkpatrick05}
{Kirkpatrick}, J.~D. 2005, \araa, 43, 195

\bibitem[{{Kirkpatrick} {et~al.}(2011){Kirkpatrick}, {Cushing}, {Gelino},
  {Griffith}, {Skrutskie}, {Marsh}, {Wright}, {Mainzer}, {Eisenhardt},
  {McLean}, {Thompson}, {Bauer}, {Benford}, {Bridge}, {Lake}, {Petty},
  {Stanford}, {Tsai}, {Bailey}, {Beichman}, {Bloom}, {Bochanski}, {Burgasser},
  {Capak}, {Cruz}, {Hinz}, {Kartaltepe}, {Knox}, {Manohar}, {Masters},
  {Morales-Calder{\'o}n}, {Prato}, {Rodigas}, {Salvato}, {Schurr}, {Scoville},
  {Simcoe}, {Stapelfeldt}, {Stern}, {Stock}, \& {Vacca}}]{kirkpatrick11}
{Kirkpatrick}, J.~D., {Cushing}, M.~C., {Gelino}, C.~R., {et~al.} 2011, \apjs,
  197, 19

\bibitem[{{Kirkpatrick} {et~al.}(1997){Kirkpatrick}, {Henry}, \&
  {Irwin}}]{kirkpatrick97}
{Kirkpatrick}, J.~D., {Henry}, T.~J., \& {Irwin}, M.~J. 1997, \aj, 113, 1421

\bibitem[{{Kirkpatrick} {et~al.}(2016){Kirkpatrick}, {Kellogg}, {Schneider},
  {Fajardo-Acosta}, {Cushing}, {Greco}, {Mace}, {Gelino}, {Wright},
  {Eisenhardt}, {Stern}, {Faherty}, {Sheppard}, {Lansbury}, {Logsdon},
  {Martin}, {McLean}, {Schurr}, {Cutri}, \& {Conrow}}]{kirkpatrick16}
{Kirkpatrick}, J.~D., {Kellogg}, K., {Schneider}, A.~C., {et~al.} 2016, \apjs,
  224, 36

\bibitem[{{Kopparapu} {et~al.}(2013){Kopparapu}, {Ramirez}, {Kasting}, {Eymet},
  {Robinson}, {Mahadevan}, {Terrien}, {Domagal-Goldman}, {Meadows}, \&
  {Deshpande}}]{kopparapu13}
{Kopparapu}, R.~K., {Ramirez}, R., {Kasting}, J.~F., {et~al.} 2013, \apj, 765,
  131

\bibitem[{{Kostov} {et~al.}(2019){Kostov}, {Schlieder}, {Barclay}, {Quintana},
  {Col{\'o}n}, {Brand e}, {Collins}, {Feinstein}, {Hadden}, {Kane},
  {Kreidberg}, {Kruse}, {Lam}, {Matthews}, {Montet}, {Pozuelos}, {Stassun},
  {Winters}, {Ricker}, {Vanderspek}, {Latham}, {Seager}, {Winn}, {Jenkins},
  {Afanasev}, {Armstrong}, {Arney}, {Boyd}, {Barentsen}, {Barkaoui}, {Batalha},
  {Beichman}, {Bayliss}, {Burke}, {Burdanov}, {Cacciapuoti}, {Carson},
  {Charbonneau}, {Christiansen}, {Ciardi}, {Clampin}, {Collins}, {Conti},
  {Coughlin}, {Covone}, {Crossfield}, {Delrez}, {Domagal-Goldman}, {Dressing},
  {Ducrot}, {Essack}, {Everett}, {Fauchez}, {Foreman-Mackey}, {Gan}, {Gilbert},
  {Gillon}, {Gonzales}, {Hamann}, {Hedges}, {Hocutt}, {Hoffman}, {Horch},
  {Horne}, {Howell}, {Hynes}, {Ireland }, {Irwin}, {Isopi}, {Jensen}, {Jehin},
  {Kaltenegger}, {Kielkopf}, {Kopparapu}, {Lewis}, {Lopez}, {Lissauer}, {Mann},
  {Mallia}, {Mandell}, {Matson}, {Mazeh}, {Monsue}, {Moran}, {Moran}, {Morley},
  {Morris}, {Muirhead}, {Mukai}, {Mullally}, {Mullally}, {Murray}, {Narita},
  {Palle}, {Pidhorodetska}, {Quinn}, {Relles}, {Rinehart}, {Ritsko},
  {Rodriguez}, {Rowden}, {Rowe}, {Sebastian}, {Sefako}, {Shahaf}, {Shporer},
  {Ta{\~n}{\'o}n Reyes}, {Tenenbaum}, {Ting}, {Twicken}, {van Belle}, {Vega},
  {Volosin}, {Walkowicz}, \& {Youngblood}}]{kostov2019}
{Kostov}, V.~B., {Schlieder}, J.~E., {Barclay}, T., {et~al.} 2019, \aj, 158, 32

\bibitem[{Lienhard(2020)}]{lienhard20}
Lienhard, F. e.~a. 2020, Monthly Notices of the Royal Astronomical Society,
  submitted

\bibitem[{Lightkurve~Collaboration {et~al.}(2018)Lightkurve~Collaboration,
  Cardoso, Hedges, Gully-Santiago, Saunders, Cody, Barclay, Hall, Sagear,
  Turtelboom, Zhang, Tzanidakis, Mighell, Coughlin, Bell, Berta-Thompson,
  Williams, Dotson, Barentsen, Collaboration, Cardoso, Hedges, Gully-Santiago,
  Saunders, Cody, Barclay, Hall, Sagear, Turtelboom, Zhang, Tzanidakis,
  Mighell, Coughlin, Bell, Berta-Thompson, Williams, Dotson, \&
  Barentsen}]{lightkurve2018}
Lightkurve~Collaboration, L., Cardoso, J. V. d.~M., Hedges, C., {et~al.} 2018,
  ascl, ascl:1812.013

\bibitem[{{Lissauer}(2007)}]{lissauer07}
{Lissauer}, J.~J. 2007, \apjl, 660, L149

\bibitem[{Luger {et~al.}(2017)Luger, Sestovic, Kruse, Grimm, Demory, Agol,
  Bolmont, Fabrycky, Fernandes, Grootel, Burgasser, Gillon, Ingalls, Jehin,
  Raymond, Selsis, Triaud, Barclay, Barentsen, Howell, Delrez, de~Wit,
  Foreman-Mackey, Holdsworth, Leconte, Lederer, Turbet, Almleaky, Benkhaldoun,
  Magain, Morris, Heng, \& Queloz}]{Luger2017a}
Luger, R., Sestovic, M., Kruse, E., {et~al.} 2017, Nature Astronomy, 1

\bibitem[{{Luhman}(2013)}]{Luhman2013}
{Luhman}, K.~L. 2013, The Astrophysical Journal Letters, 767, L1

\bibitem[{{Lustig-Yaeger} {et~al.}(2019){Lustig-Yaeger}, {Meadows}, \&
  {Lincowski}}]{LustigYaeger2019}
{Lustig-Yaeger}, J., {Meadows}, V.~S., \& {Lincowski}, A.~P. 2019, The
  Astronomical Journal, 158, 27

\bibitem[{{Macdonald} \& {Cowan}(2019)}]{macdonald19}
{Macdonald}, E. J.~R. \& {Cowan}, N.~B. 2019, arXiv e-prints, arXiv:1908.10873

\bibitem[{{Madhusudhan}(2019)}]{Madhusudhan19}
{Madhusudhan}, N. 2019, \araa, 57, 617

\bibitem[{{Mann} {et~al.}(2019){Mann}, {Dupuy}, {Kraus}, {Gaidos}, {Ansdell},
  {Ireland}, {Rizzuto}, {Hung}, {Dittmann}, {Factor}, {Feiden}, {Martinez},
  {Ru{\'\i}z-Rodr{\'\i}guez}, \& {Thao}}]{Mann2019}
{Mann}, A.~W., {Dupuy}, T., {Kraus}, A.~L., {et~al.} 2019, The Astrophysical
  Journal, 871, 63

\bibitem[{{Menou}(2013)}]{menou13}
{Menou}, K. 2013, \apj, 774, 51

\bibitem[{{Ment} {et~al.}(2019){Ment}, {Dittmann}, {Astudillo-Defru},
  {Charbonneau}, {Irwin}, {Bonfils}, {Murgas}, {Almenara}, {Forveille}, {Agol},
  {Ballard}, {Berta-Thompson}, {Bouchy}, {Cloutier}, {Delfosse}, {Doyon},
  {Dressing}, {Esquerdo}, {Haywood}, {Kipping}, {Latham}, {Lovis}, {Newton},
  {Pepe}, {Rodriguez}, {Santos}, {Tan}, {Udry}, {Winters}, \&
  {W{\"u}nsche}}]{ment2019}
{Ment}, K., {Dittmann}, J.~A., {Astudillo-Defru}, N., {et~al.} 2019, \aj, 157,
  32

\bibitem[{{Miguel} {et~al.}(2020){Miguel}, {Cridland}, {Ormel}, {Fortney}, \&
  {Ida}}]{miguel20}
{Miguel}, Y., {Cridland}, A., {Ormel}, C.~W., {Fortney}, J.~J., \& {Ida}, S.
  2020, \mnras, 491, 1998

\bibitem[{{Montgomery} \& {Laughlin}(2009)}]{montgomery09}
{Montgomery}, R. \& {Laughlin}, G. 2009, \icarus, 202, 1

\bibitem[{{Morley} {et~al.}(2017){Morley}, {Kreidberg}, {Rustamkulov},
  {Robinson}, \& {Fortney}}]{Morley2017}
{Morley}, C.~V., {Kreidberg}, L., {Rustamkulov}, Z., {Robinson}, T., \&
  {Fortney}, J.~J. 2017, The Astrophysical Journal, 850, 121

\bibitem[{{Morris} {et~al.}(2018){Morris}, {Tollerud}, {Sip{\H o}cz}, {Deil},
  {Douglas}, {Berlanga Medina}, {Vyhmeister}, {Smith}, {Littlefair},
  {Price-Whelan}, {Gee}, \& {Jeschke}}]{astroplan2018}
{Morris}, B.~M., {Tollerud}, E., {Sip{\H o}cz}, B., {et~al.} 2018, \aj, 155,
  128

\bibitem[{Muirhead {et~al.}(2012)Muirhead, Johnson, Apps, Carter, Morton,
  Fabrycky, Pineda, Bottom, Rojas-Ayala, Schlawin, Hamren, Covey, Crepp,
  Stassun, Pepper, Hebb, Kirby, Howard, Isaacson, Marcy, Levitan, Diaz-Santos,
  Armus, \& Lloyd}]{Muirhead12}
Muirhead, P.~S., Johnson, J.~A., Apps, K., {et~al.} 2012, The Astrophysical
  Journal, 747, 144

\bibitem[{{Murray} {et~al.}(2020){Murray}, {Delrez}, {Pedersen}, {Queloz},
  {Gillon}, {Burdanov}, {Ducrot}, {Garcia}, {Lienhard}, {Demory}, {Jehin},
  {McCormac}, {Sebastian}, {Sohy}, {Thompson}, {Triaud}, {Van Grootel},
  {G{\"u}nther}, \& {Huang}}]{murray20}
{Murray}, C.~A., {Delrez}, L., {Pedersen}, P.~P., {et~al.} 2020, \mnras, 495,
  2446

\bibitem[{{Nowak} {et~al.}(2020){Nowak}, {Luque}, {Parviainen}, {Pall{\'e}},
  {Molaverdikhani}, {B{\'e}jar}, {Lillo-Box}, {Rodr{\'\i}guez-L{\'o}pez},
  {Caballero}, {Zechmeister}, {Passegger}, {Cifuentes}, {Schweitzer}, {Narita},
  {Cale}, {Espinoza}, {Murgas}, {Zapatero Osorio}, {Pozuelos}, {Aceituno},
  {Amado}, {Barkaoui}, {Barrado}, {Bauer}, {Benkhaldoun}, {Caldwell},
  {Casasayas Barris}, {Chaturvedi}, {Chen}, {Collins}, {Collins},
  {Cort{\'e}s-Contreras}, {Crossfield}, {de Le{\'o}n}, {D{\'\i}ez Alonso},
  {Dreizler}, {El Mufti}, {Esparza-Borges}, {Essack}, {Fukui}, {Gillon},
  {Guerra}, {Hatzes}, {Henning}, {Herrero}, {Hesse}, {Hirano}, {Howell},
  {Jeffers}, {Jehin}, {Jenkins}, {Kaminski}, {Kemmer}, {Kielkopf},
  {Kossakowski}, {Kotani}, {K{\"u}rster}, {Lafarga}, {Latham}, {Law},
  {Lissauer}, {Lodieu}, {Madrigal-Aguado}, {Mann}, {Massey}, {Matson},
  {Matthews}, {Monta{\~n}{\'e}s-Rodr{\'\i}guez}, {Montes}, {Morales}, {Mori},
  {Nagel}, {Oshagh}, {Pedraz}, {Plavchan}, {Pollacco}, {Quirrenbach},
  {Reffert}, {Reiners}, {Ribas}, {Rose}, {Schlecker}, {Schlieder}, {Seager},
  {Stangret}, {Stock}, {Tamura}, {Teske}, {Trifonov}, {Twicken}, {Watanabe},
  {Wittrock}, {Ziegler}, \& {Zohrabi}}]{nowak2020}
{Nowak}, G., {Luque}, R., {Parviainen}, H., {et~al.} 2020, arXiv e-prints,
  arXiv:2003.01140

\bibitem[{{Nutzman} \& {Charbonneau}(2008)}]{Nutzman2008}
{Nutzman}, P. \& {Charbonneau}, D. 2008, Publications of the Astronomical
  Society of the Pacific, 120, 317

\bibitem[{{Payne} \& {Lodato}(2007)}]{payne07}
{Payne}, M.~J. \& {Lodato}, G. 2007, \mnras, 381, 1597

\bibitem[{{Pecaut} \& {Mamajek}(2013)}]{PM2013}
{Pecaut}, M.~J. \& {Mamajek}, E.~E. 2013, \apjs, 208, 9

\bibitem[{{Pozuelos} {et~al.}(2020){Pozuelos}, {Su{\'a}rez}, {de El{\'\i}a},
  {Berdi{\~n}as}, {Bonfanti}, {Dugaro}, {Gillon}, {Jehin}, {G{\"u}nther}, {Van
  Grootel}, {Garcia}, {Thuillier}, {Delrez}, \& {Rod{\'o}n}}]{pozuelos2020}
{Pozuelos}, F.~J., {Su{\'a}rez}, J.~C., {de El{\'\i}a}, G.~C., {et~al.} 2020,
  arXiv e-prints, arXiv:2006.09403

\bibitem[{{Quinn} {et~al.}(2019){Quinn}, {Becker}, {Rodriguez}, {Hadden},
  {Huang}, {Morton}, {Adams}, {Armstrong}, {Eastman}, {Horner}, {Kane},
  {Lissauer}, {Twicken}, {Vanderburg}, {Wittenmyer}, {Ricker}, {Vanderspek},
  {Latham}, {Seager}, {Winn}, {Jenkins}, {Agol}, {Barkaoui}, {Beichman},
  {Bouchy}, {Bouma}, {Burdanov}, {Campbell}, {Carlino}, {Cartwright},
  {Charbonneau}, {Christiansen}, {Ciardi}, {Collins}, {Collins}, {Conti},
  {Crossfield}, {Daylan}, {Dittmann}, {Doty}, {Dragomir}, {Ducrot}, {Gillon},
  {Glidden}, {Goeke}, {Gonzales}, {He{\l}miniak}, {Horch}, {Howell}, {Jehin},
  {Jensen}, {Kielkopf}, {Kristiansen}, {Law}, {Mann}, {Marmier}, {Matson},
  {Matthews}, {Mazeh}, {Mori}, {Murgas}, {Murray}, {Narita}, {Nielsen},
  {Ottoni}, {Palle}, {Paw{\l}aszek}, {Pepe}, {Pitogo de Leon}, {Pozuelos},
  {Relles}, {Schlieder}, {Sebastian}, {S{\'e}gransan}, {Shporer}, {Stassun},
  {Tamura}, {Udry}, {Waite}, {Winters}, \& {Ziegler}}]{quinn2019}
{Quinn}, S.~N., {Becker}, J.~C., {Rodriguez}, J.~E., {et~al.} 2019, \aj, 158,
  177

\bibitem[{{Raymond}(2007)}]{raymond07}
{Raymond}, S. 2007, in American Astronomical Society Meeting Abstracts, Vol.
  210, American Astronomical Society Meeting Abstracts \#210, 110.03

\bibitem[{{Reyl{\'e}}(2018)}]{reyle18}
{Reyl{\'e}}, C. 2018, \aap, 619, L8

\bibitem[{{Ricker} {et~al.}(2015){Ricker}, {Winn}, {Vanderspek}, {Latham},
  {Bakos}, {Bean}, {Berta-Thompson}, {Brown}, {Buchhave}, {Butler}, {Butler},
  {Chaplin}, {Charbonneau}, {Christensen-Dalsgaard}, {Clampin}, {Deming},
  {Doty}, {De Lee}, {Dressing}, {Dunham}, {Endl}, {Fressin}, {Ge}, {Henning},
  {Holman}, {Howard}, {Ida}, {Jenkins}, {Jernigan}, {Johnson}, {Kaltenegger},
  {Kawai}, {Kjeldsen}, {Laughlin}, {Levine}, {Lin}, {Lissauer}, {MacQueen},
  {Marcy}, {McCullough}, {Morton}, {Narita}, {Paegert}, {Palle}, {Pepe},
  {Pepper}, {Quirrenbach}, {Rinehart}, {Sasselov}, {Sato}, {Seager},
  {Sozzetti}, {Stassun}, {Sullivan}, {Szentgyorgyi}, {Torres}, {Udry}, \&
  {Villasenor}}]{Ricker2015}
{Ricker}, G.~R., {Winn}, J.~N., {Vanderspek}, R., {et~al.} 2015, Journal of
  Astronomical Telescopes, Instruments, and Systems, 1, 014003

\bibitem[{{Sagear} {et~al.}(2019){Sagear}, {Skinner}, \& {Muirhead}}]{segear19}
{Sagear}, S.~A., {Skinner}, J.~N., \& {Muirhead}, P.~S. 2019, arXiv e-prints,
  arXiv:1912.04286

\bibitem[{{Scholz}(2014)}]{scholz142}
{Scholz}, R.~D. 2014, \aap, 561, A113

\bibitem[{{Scholz}(2020)}]{scholz20}
{Scholz}, R.-D. 2020, arXiv e-prints, arXiv:2003.10949

\bibitem[{{Schoonenberg} {et~al.}(2019){Schoonenberg}, {Liu}, {Ormel}, \&
  {Dorn}}]{schoonenberg19}
{Schoonenberg}, D., {Liu}, B., {Ormel}, C.~W., \& {Dorn}, C. 2019, \aap, 627,
  A149

\bibitem[{{Skrutskie} {et~al.}(2006){Skrutskie}, {Cutri}, {Stiening},
  {Weinberg}, {Schneider}, {Carpenter}, {Beichman}, {Capps}, {Chester},
  {Elias}, {Huchra}, {Liebert}, {Lonsdale}, {Monet}, {Price}, {Seitzer},
  {Jarrett}, {Kirkpatrick}, {Gizis}, {Howard}, {Evans}, {Fowler}, {Fullmer},
  {Hurt}, {Light}, {Kopan}, {Marsh}, {McCallon}, {Tam}, {Van Dyk}, \&
  {Wheelock}}]{2MASS}
{Skrutskie}, M.~F., {Cutri}, R.~M., {Stiening}, R., {et~al.} 2006, The
  Astronomical Journal, 131, 1163

\bibitem[{{Smart} {et~al.}(2019){Smart}, {Marocco}, {Sarro}, {Barrado},
  {Beamin}, {Caballero}, \& {Jones}}]{smart19}
{Smart}, R.~L., {Marocco}, F., {Sarro}, L.~M., {et~al.} 2019, arXiv e-prints,
  arXiv:1902.07571

\bibitem[{{Stassun} \& {Torres}(2018)}]{Stassun2018}
{Stassun}, K.~G. \& {Torres}, G. 2018, \apj, 862, 61

\bibitem[{{Sullivan} {et~al.}(2015){Sullivan}, {Winn}, {Berta-Thompson},
  {Charbonneau}, {Deming}, {Dressing}, {Latham}, {Levine}, {McCullough},
  {Morton}, {Ricker}, {Vanderspek}, \& {Woods}}]{Sullivan2015}
{Sullivan}, P.~W., {Winn}, J.~N., {Berta-Thompson}, Z.~K., {et~al.} 2015, The
  Astrophysical Journal, 809, 77

\bibitem[{{Tamburo} \& {Muirhead}(2019)}]{tamburo19}
{Tamburo}, P. \& {Muirhead}, P.~S. 2019, \pasp, 131, 114401

\bibitem[{{Triaud} {et~al.}(2020){Triaud}, {Burgasser}, {Burdanov}, {Kunovac
  Hod{\v{z}}i{\'c}}, {Alonso}, {Bardalez Gagliuffi}, {Delrez}, {Demory}, {de
  Wit}, {Ducrot}, {Hessman}, {Husser}, {Jehin}, {Pedersen}, {Queloz},
  {McCormac}, {Murray}, {Sebastian}, {Thompson}, {Van Grootel}, \&
  {Gillon}}]{triaud20}
{Triaud}, A. H.~M.~J., {Burgasser}, A.~J., {Burdanov}, A., {et~al.} 2020,
  Nature Astronomy [\eprint[arXiv]{2001.07175}]

\bibitem[{{Tsiaras} {et~al.}(2019){Tsiaras}, {Waldmann}, {Tinetti}, {Tennyson},
  \& {Yurchenko}}]{tsiaras2019}
{Tsiaras}, A., {Waldmann}, I.~P., {Tinetti}, G., {Tennyson}, J., \&
  {Yurchenko}, S.~N. 2019, Nature Astronomy, 3, 1156

\bibitem[{{Vanderspek} {et~al.}(2019){Vanderspek}, {Huang}, {Vanderburg},
  {Ricker}, {Latham}, {Seager}, {Winn}, {Jenkins}, {Burt}, {Dittmann},
  {Newton}, {Quinn}, {Shporer}, {Charbonneau}, {Irwin}, {Ment}, {Winters},
  {Collins}, {Evans}, {Gan}, {Hart}, {Jensen}, {Kielkopf}, {Mao}, {Waalkes},
  {Bouchy}, {Marmier}, {Nielsen}, {Ottoni}, {Pepe}, {S{\'e}gransan}, {Udry},
  {Henry}, {Paredes}, {James}, {Hinojosa}, {Silverstein}, {Palle},
  {Berta-Thompson}, {Crossfield}, {Davies}, {Dragomir}, {Fausnaugh}, {Glidden},
  {Pepper}, {Morgan}, {Rose}, {Twicken}, {Villase{\~n}or}, {Yu}, {Bakos},
  {Bean}, {Buchhave}, {Christensen-Dalsgaard}, {Christiansen}, {Ciardi},
  {Clampin}, {De Lee}, {Deming}, {Doty}, {Jernigan}, {Kaltenegger}, {Lissauer},
  {McCullough}, {Narita}, {Paegert}, {Pal}, {Rinehart}, {Sasselov}, {Sato},
  {Sozzetti}, {Stassun}, \& {Torres}}]{Vanderspek2019}
{Vanderspek}, R., {Huang}, C.~X., {Vanderburg}, A., {et~al.} 2019, The
  Astrophysical Journal Letters, 871, L24

\bibitem[{{Wenger} {et~al.}(2000){Wenger}, {Ochsenbein}, {Egret}, {Dubois},
  {Bonnarel}, {Borde}, {Genova}, {Jasniewicz}, {Lalo{\"e}}, {Lesteven}, \&
  {Monier}}]{wenger00}
{Wenger}, M., {Ochsenbein}, F., {Egret}, D., {et~al.} 2000, \aaps, 143, 9

\bibitem[{{Winn}(2010)}]{Winn2010}
{Winn}, J.~N. 2010, {Exoplanet Transits and Occultations} (University of
  Arizona Press), 55 -- 77

\end{thebibliography}


\end{document}